\documentclass{aims} % Use the aims.cls file to compile your paper
\usepackage{amsmath}
\usepackage{paralist}
\usepackage[misc]{ifsym}
\usepackage{epsfig} %For pictures: screened artwork should be set up with an 85 or 100 line screen
\usepackage{epstopdf} %This is to transfer .eps figure to .pdf figure; please compile your article using PDFLaTex or PDFTeXify.
\usepackage[colorlinks=true]{hyperref}
\hypersetup{urlcolor=blue, citecolor=red}
\allowdisplaybreaks

\def\currentvolume{X}
\def\currentissue{X}
\def\currentyear{200X}
\def\currentmonth{XX}
\def\ppages{X-XX}
\def\DOI{10.3934/xx.xxxxxxx}

\usepackage[style=ieee, sorting=nyt, citestyle=numeric-comp]{biblatex}

\usepackage[nolist]{acronym}
\usepackage[ruled]{algorithm2e}
\usepackage{amssymb}
\usepackage{mathtools}
\usepackage{dsfont}
\usepackage[format=plain]{caption}
\usepackage{subcaption}
\usepackage[makeroom]{cancel}
\usepackage{tikz}
\tikzstyle{compartment3} = [rectangle, minimum width=.8cm, minimum height=.8cm, text centered, draw=black, fill=lightgray]
\tikzstyle{plate} = [rectangle, minimum width=6.3cm, minimum height=4cm, text centered, draw=black, fill=white]
\tikzstyle{determ} =[thick, dashed, ->, gray]
\tikzstyle{node} = [circle, minimum width=.8cm, minimum height=.8cm, text centered, draw=black, fill=white]
\tikzstyle{nodeknown} = [circle, minimum width=0.8cm, minimum height=0.8cm, text centered, draw=black, fill=lightgray]
\tikzstyle{namecompartment} = [rectangle, minimum width=0.5cm, minimum height=0.1cm, text centered]

\newtheorem{theorem}{Theorem}[section]

\newtheorem{lemma}[theorem]{Lemma}

\theoremstyle{definition}

\title[The Lifebelt Particle Filter]
{The Lifebelt Particle Filter for robust estimation from low-valued count data} % Only the first word and proper nouns should be capitalized.

\author[Corbella et al.]{}

\subjclass{Primary: 62M20; Secondary: 60G35.}
\keywords{discrete system, low counts, particle collapse, deterministic mixtures, lifebelt particle filter.}

\thanks{The first author is supported by [Bayes4Health EPSRC Grant EP/R018561/1]}

\thanks{$^*$Corresponding author: Alice Corbella}

\begin{document}
\maketitle

% Enter the first author's name and email address; email addresses are required for each author.
% Use footnote notations to indicate address and affiliations with commas between numbers if more than one address applies; see below for examples.
\centerline{\scshape
Alice Corbella$^{{\href{mailto:alice.corbella@warwick.ac.uk}{\textrm{\Letter}}}*1}$,
Trevelyan J. McKinley$^{{\href{mailto:t.mckinley@exeter.ac.uk}{\textrm{\Letter}}}*2}$,
Paul J. Birrell$^{{\href{mailto:paul.birrel@ukhsa.gov.uk}{\textrm{\Letter}}}*3,4}$,}
\centerline{\scshape
Daniela De Angelis$^{{\href{mailto:daniela.deangelis@mrc-bsu.cam.ac.uk}{\textrm{\Letter}}}*3,4}$,
Anne M. Presanis$^{{\href{mailto:anne.presanis@mrc-bsu.cam.ac.uk}{\textrm{\Letter}}}*3}$,
Gareth O. Roberts$^{{\href{mailto:g.o.roberts@warwick.ac.uk}{\textrm{\Letter}}}*1}$,}
\centerline{\scshape
and Simon E. F. Spencer$^{{\href{mailto:s.e.f.spencer@warwick.ac.uk}{\textrm{\Letter}}}1}$}
\medskip

{\footnotesize
% Enter the full affiliation and country name:
% Do not insert commas or periods at the end of lines.
\centerline{$^1$Department of Statistics, University of Warwick (UK)}
} % Do not forget to end {\footnotesize with the sign }

\medskip

{\footnotesize
% Enter the full affiliation and country name:
\centerline{$^2$University of Exeter Medical School, University of Exeter (UK)}
}
\medskip

{\footnotesize
% Enter the full affiliation and country name:
\centerline{$^3$MRC Biostatistics Unit, University of Cambridge (UK)}
}\medskip

{\footnotesize
% Enter the full affiliation and country name:
\centerline{$^4$UK Health Security Agency (UK)}
}
\bigskip

% The name of the handling editor will be entered by AIMS production staff.
% "Communicated by Handling Editor" is not needed for special issue.
\centerline{(Communicated by Handling Editor)}

%%%%%%%%%%%%%%%%%%%%%%%%%%%%%%%%%%%%%%%%%%%%%%%%%%%%%%%
%             5. ABSTRACT
%%%%%%%%%%%%%%%%%%%%%%%%%%%%%%%%%%%%%%%%%%%%%%%%%%%%%%%

\begin{abstract}
Particle filtering methods can be applied to {estimation} problems in discrete spaces on bounded domains, to sample from and marginalise over unknown hidden states. As in continuous settings, problems such as particle degradation can arise: proposed particles {can be incompatible with the data, lying in low probability regions or outside the boundary constraints,} and the discrete system could result in all particles having weights of zero. 

In this paper we introduce the Lifebelt Particle Filter (LBPF), a novel method for robust likelihood estimation in low-valued count problems. The LBPF combines a standard particle filter with one (or more) \textit{lifebelt particles} which, by construction, {lie within the boundaries of the discrete random variables, and therefore are compatible with the data}. A mixture of resampled and non-resampled particles allows for the preservation of the lifebelt particle, which, together with the remaining particle swarm, provides samples from the filtering distribution, and can be used to generate {unbiased} estimates of the likelihood. 

{The main benefit of the LBPF is that only one or few, wisely chosen, particles are sufficient to prevent particle collapse. Differently from other methods, there is no need to increase the number of particles, and therefore the computational effort, in regions of the parameter space that generate less likely hidden states. }

The LBPF can be used within a pseudo-marginal scheme to draw inferences on static parameters, $ \boldsymbol{\theta} $, governing the system. We address here the estimation of a parameter governing probabilities of deaths and recoveries of hospitalised patients during an epidemic.

\end{abstract}

%%%%%%%%%%%%%%%%%%%%%%%%%%%%%%%%%%%%%%%%%%%%%%%%%%%%%%
%                   6. BODY
%%%%%%%%%%%%%%%%%%%%%%%%%%%%%%%%%%%%%%%%%%%%%%%%%%%%%%

\section{Introduction}
\Acp{SSM} are stochastic processes that make use of a latent variable representation to describe a dynamical phenomenon \cite{schon2018probabilistic}. \acp{SSM} are used in disparate fields, from object positioning in engineering \cite{gilks2001following, solin2018modeling}, to evolution of weather conditions \cite{anderson1996method}, to virus spread in a population \cite{breto2009time, dukic2012tracking}.

A \ac{SSM} uses a state system to describe the unknown hidden dynamics, of which only noisy and partial observations are available. This system can be explored via \ac{SMC}, i.e. iterative use of simulation algorithms that sample from the hidden states and use the available observations to infer the parameters governing the system. These methods, as the proposals of this paper, were developed mainly within the Bayesian framework. For a complete review of \ac{SMC} methods see a recent volume by Chopin and Papaspiliopoulos \cite{chopin2020introduction} and key papers \cite{arulampalam2002tutorial, doucet2009tutorial, schon2018probabilistic}.

While most of the examples presented in the \ac{SMC} literature consider continuous unbounded hidden states, many real life dynamical systems involve counts, requiring discrete and bounded \acp{SSM} (e.g. epidemic models, queues, change-points models). The application of standard \ac{SMC} methods in these settings is often challenging, and problem-specific proposals are formulated to sample efficiently from the hidden states \cite[e.g., ][]{mckinley2020efficient, whiteley2021inference}. 

In this paper we propose the \ac{LBPF} a new \ac{SMC} algorithm to sample from the bounded hidden state of a discrete-valued \ac{SSM}. The algorithm directs specific simulations to the areas of the hidden state within the boundaries by using deterministic proposals and a newly-formulated resampling scheme. The combination of these two elements allows the algorithm to always retain a simulated trajectory within the boundaries.  

Section \ref{s2} introduces the background and states the problem under analysis. In this section an important motivating example is also presented, where a \ac{SSM} is used to describe the generation of data on cases and death by a specific disease; standard \acp{SMC} performs poorly in this context due to the strict boundaries that constrain the hidden state. 
Section \ref{s3} outlines the \ac{LBPF} defining the specification of the algorithm. 
Section \ref{s4} returns to the motivating example presented in Section \ref{s2}, and highlights why standard \ac{SMC} algorithms perform poorly, showing how the \ac{LBPF} provides a suitable alternative. Several versions of the \ac{LBPF} are presented in this section, showing the flexibility of the algorithm. 
Section \ref{s5} benchmarks the proposed algorithm against an alternative solution and Section \ref{s6} draws conclusions and discusses future work. 

\section{Background and Problem statement}\label{s2}
Let us consider a discrete-time Markovian \ac{SSM} defined by $\{ X_t\}_{t=1}^{T} $, the unknown state process (or hidden process), and $\{ Y_t\}_{t=1}^{T} $, the observational process (or emission process). 
Denote with $\Omega_{X_t}$ the support of $X_t$ and $\Omega_{Y_t}$ the support of $Y_t$ and assume the process is discrete-valued, i.e. $\Omega_{X_t}\subseteq \mathbb{N} $ and $ \Omega_{Y_t}\subseteq\mathbb{N} $ for $ t=0,1, \dots T$. Let $\boldsymbol{\theta}\in \mathbb{R}^d$ be the static parameter governing the \ac{SSM} so that the probability of the state process is: 
\begin{equation}\label{eq:state}
p(x_{0:T}|\boldsymbol{\theta})=	p(x_0|\boldsymbol{\theta})\prod_{t=1}^{T}p(x_t|x_{t-1}, \boldsymbol{\theta})
\end{equation}and the emission/observation probability is: 
\begin{equation}\label{eq:obs}
p(y_t|x_t, \boldsymbol{\theta})
\end{equation}
so that the joint distribution can be factorised as follows:
\begin{equation}\label{eq:fullmod}
p(y_t, x_t, \boldsymbol{\theta}) = p(\boldsymbol{\theta})p(x_0|\boldsymbol{\theta})\prod_{t=1}^{T}p(x_t|x_{t-1}, \boldsymbol{\theta})p(y_t|x_t, \boldsymbol{\theta})
\end{equation}
with $p(\boldsymbol{\theta})$ denoting the prior information on the static parameter.

The likelihood $p(y_{1:T}|\boldsymbol{\theta})$, which is key to performing inference on $\boldsymbol{\theta}$, involves the marginalisation over all the values of the hidden states $X_{1:T}$, making the computation analytically intractable. In \ac{SMC} this problem is addressed by approximating $p(y_{1:T}|\boldsymbol{\theta})$ using samples from the filtering distribution $p(x_{0:t}|y_{1:t})$, iteratively for $t=1, \dots, T$ as briefly outlined below. 

Let $\left\lbrace x_{t-1}^{(n)}, {{w}}_{t-1}^{(n)}\right\rbrace_{n=1}^N $ be a weighted sample of size $N$ from the filtering distribution at time $t-1$:
\begin{equation}\label{eq:diracapp}
\sum_{n=1}^{N}\delta_{x_{t-1}^{(n)}}(x )\widetilde{w}_{t-1}^{(n)} \approx p(X_{t-1}=x|y_{1:t-1}, \boldsymbol{\theta}) \qquad \forall x \in \Omega_{X_{t-1}}
\end{equation} 
where $\delta_{x_{t-1}^{(n)}}(x )$ is the unit delta measure concentrated in $x_{t-1}^{(n)}$, and $\widetilde{w}_{t-1}^{(n)}$ for $n=1, \dots N$ are self-normalised weights such that: $\widetilde{w}_{t-1}^{(n)}=\frac{{w}_{t-1}^{(n)}}{\sum_m{w}_{t-1}^{(m)}}$ and $\sum_{n=1}^{N}\widetilde{w}_{t-1}^{(n)} =1 $. These weighted samples are called particles. A weighed sample from the filtering distribution at time $t$ can be obtained by the following steps which constitute the \ac{SMC} scheme named \ac{SIRS}.

[Step 1] A  {resampling step} can be used to obtain an equally-weighted sample. Denote with $\mathcal{C}\left\lbrace \boldsymbol{c}, \boldsymbol{w}\right\rbrace $ a categorical \ac{r.v.} with values $\boldsymbol{c}$ and weights $\boldsymbol{w}$. A set of indices can be sampled:
\begin{equation}\label{eq:resample}
a_j\sim \mathcal{C} \left\lbrace 1, 2, \dots , N ; \widetilde{w}_{t-1}^{(1)}, \widetilde{w}_{t-1}^{(2)}, \dots \widetilde{w}_{t-1}^{(N)} \right\rbrace  \qquad\text{ for }j=1,2, \dots N
\end{equation}
to obtain the sample $\left\lbrace x_{t-1}^{(a_j)}, \frac{1}{N} \right\rbrace_{j=1}^N$.

[Step 2] An importance distribution, denoted by $q(\cdot)$, can then be used to draw proposed samples from the hidden state at time $t$, usually conditionally on the sample available at time $ t-1$:
\begin{equation}\label{eq:prop}
x_{t}^{(n)} \sim q(\cdot|x_{t-1}) \qquad \text{ for } n=1, 2,  \dots N.
\end{equation}

[Step 3] Importance weights can be assigned to each particle according to the joint distribution of the hidden state $x_t$ and the observation $y_t$:
\begin{equation}\label{eq:weights}
w_{t}^{(n)}=\frac{p(y_t, x_t^{(n)}|x_{t-1}, \boldsymbol{\theta})}{q(x_t^{(n)}|x_{t-1})}\qquad \text{ for } n=1, 2,  \dots N.
\end{equation}

{Note that [Step 1] of the scheme is optional: resampling the particles allows for more homogeneous, in fact equal, weights. If [Step 1] is omitted, then the algorithm is simply called \ac{SIS} and the weights in [Step 3] can be modified accordingly: 
\begin{equation*}\label{eq:weightssis}
w_{t}^{(n)}=\frac{p(y_t, x_t^{(n)}|x_{t-1}, \boldsymbol{\theta})}{q(x_t^{(n)}|x_{t-1})}	w_{t-1}^{(n)}.
\end{equation*}
Another notable scheme is the \acl{BPF}\acused{BPF} \cite[BPF; ][]{gordon1993novel}, which simplifies the weights by choosing as importance distribution the state distribution: $q(x_t|x_{t-1})=p(x_t|x_{t-1})$.

The weights drawn in equation \eqref{eq:weights} can be used to approximate the $t$-th component of the likelihood of the observations given a value of $\boldsymbol{\theta}$ as follows:
\begin{equation}\label{eq:liket}
\widehat{p}(y_{t}|y_{1:t-1}, \boldsymbol{\theta}) = \frac{1}{N} \sum_{n=1}^{n}w_{t}^{(n)},
\end{equation}
and an estimate of the value of the likelihood of the parameter $\boldsymbol{\theta}$ is :
\begin{equation}\label{eq:like1T}
\widehat{p}(y_{1:T}|\boldsymbol{\theta})=\prod_{t=1}^{T}	\widehat{p}(y_{t}|y_{1:t-1}, \boldsymbol{\theta}).
\end{equation}

Estimate \eqref{eq:like1T}}
can be embedded in a \ac{MCMC} algorithm, such as the \ac{MH} algorithm, enabling exact inference of the posterior of interest, $ p\left(\boldsymbol{\theta}|{y}_{1:T}\right) $, {through} {pseudo-marginal methods} \cite{andrieu2009pseudo}, or in another sequential method sampling from a series of distributions that converge to the target of interest \cite{chopin2013smc2}. Here we use an \ac{SMC} approximation of the likelihood within an \ac{MCMC} algorithm, so-called \ac{pMCMC} \cite{andrieu2010particle}. 

Only the essential elements of the \ac{SMC} schemes for state inference are reported above. For a complete review see the references reported in the introduction and for the proof of unbiasedness of estimator \eqref{eq:like1T} see \cite{pitt2012some}.

\subsection{{Challenges of sequential importance methods}}

{\Ac{SMC} samplers are based on importance sampling, hence their performance depends on the discrepancy between the importance distribution $q(\cdot)$ used in [Step 2] and the target distribution \cite{robert2004monte}. }
If {the importance distribution is very different from the filtering distribution, the majority of} the particles {proposed} for the latent state {will fall in a low-probability region of the} true filtering distribution. This leads to {particle degradation}: as the sequential approximation progresses, fewer and fewer particles have a non-negligible importance weight $	w_{t}^{(n)}$, and the remaining particles become non-representative of the filtering distribution. The most serious outcome is {particle collapse}, which happens when all the particle have weight equal to 0:
{\begin{equation*}
w_{t}^{(n)}=\frac{p(y_t, x_t^{(n)}|x_{t-1}, \boldsymbol{\theta})}{q(x_t^{(n)}|x_{t-1})}=0 \qquad \forall n.
\end{equation*}}

This problem is exacerbated in \acp{SSM} {where the state distributions are discrete and/or bounded, (e.g. chained systems of Binomial or Multinomial \acp{r.v.})}: the support of these variables is restricted and it can be very hard to find an importance distribution that covers the support while proposing values around the mass of the filtering distribution. 

The problem is even more severe when the observation distribution is highly concentrated (which can happen as a consequence of low value data), leaving little space for importance proposals with weight greater than 0. The {chained} dynamics of many such \acp{SSM} are the underlying cause of this problem: the particles sampled need to be coherent not only with the observations at time point $ t $, but also with the future observations, to allow the full run of the importance sampling mechanism {from $t=1$ to $T$.} Atypical data and outliers challenge even more these mechanisms, provoking full particle collapse when just one new data point is included after a fairly well-behaved particle set up to that point. 

\subsection{{Proposals to prevent particle collapse}}

Some attempts have been made to extend the basic \ac{SMC} mechanisms to be robust in discrete, low-count systems. 
{
One possibility is to use more information to formulate distribution $q(\cdot)$ in [Step 2]: the \acl{AuPF}\acused{BPF} \cite[AuPF; ][]{pitt1999filtering} exploits data at time $t$, $y_t$, to draw proposals that should be in high-probability regions of the filtering distribution. To allow for proposed particles that are compatible with all the future observations, one could use an \ac{AuPF} with importance distribution: $q(x_t|y_t, y_{t+1}, \dots, y_{T})$. Unfortunately, such distribution is likely to be very difficult to formulate, especially in complex contexts with chained temporal structures. 

Other methods work on the number of particles sampled, rather then their distribution}: the \acl{APF}\acused{APF} \cite[APF; ][]{delmoral2015alive} is a \ac{SMC} method that allows for an unbounded number of particles $ n_t $. The idea behind the algorithm is to keep sampling proposed particles until a desired number of particles $ N $ with non-zero weight is obtained. \cite{drovandi2016exact} looked at the application of the \ac{APF} in the context of low-count data and observed that the algorithm can have potentially an unbounded cost. {Even if an artificial stopping criterion is used, placing a limit on the computations executed,
the \ac{APF} is likely to take long time to run if the value of  $ \boldsymbol{\theta} $ has low likelihood (\cite{drovandi2016exact}, Remark 4) e.g. for parameter proposals towards the tails.} \cite{mckinley2020efficient} exploited much of the potential of the \ac{APF} in this context, allowing for early rejection of bad values of $ \boldsymbol{\theta} $. Despite their efforts and improvements, the authors note that the algorithm can remain substantially computationally expensive, hindering its use for longer time series or in contexts where real-time inference is needed. 

%{\color{gray}
%The \ac{NPF} $ \to $ idea is to direct the samples where the mass of the filtering distribution is.non-exact. ?Possibly non suitable for discrete problems?}

\subsection{Motivation: hospitalised case fatality risk estimation}\label{ss2.3}
{Inference during a response an emerging epidemic provide an important example of discrete bounded \ac{SSM}, requiring new \ac{SMC} routines for inference. }
Estimation of the \ac{CFR}, i.e. the probability of death given a disease, is crucial during emerging epidemics: the timely availability of estimates for the risk of death for different case-definitions (e.g., \acl{hCFR}, hCFR, \acl{sCFR}, sCFR) can provide invaluable information for a health-policy response. 
{In this context, time series on the number of cases at different levels of severity are usually collected: frequently only outcomes of higher severity (e.g. appearance of symptoms, hospitalization, death) are reported, while positive outcomes (e.g. disappearance of symptoms, recovery, discharge from hospital) are rarely monitored.}

We consider the situation when counts of confirmed cases admitted to hospital and confirmed deaths in hospital are recorded. This is the type of information available in England: a severe influenza surveillance system (USISS;  \cite{health2011sourcesa}), now replaced by the SARI Watch survey \cite{SARIWatch}, collects data on the weekly number of cases admitted to intensive care and deaths with confirmed influenza.

Let $ y_t $ and $ h_t $  denote the number of confirmed deaths and hospitalisations that happened during time interval indexed by $ t $, respectively. Estimates of the \ac{CFR} based on cumulative-counts such as
%\begin{equation}\label{key}
$
\widehat{CFR}_t=\frac{\sum_{t=1}^{T}y_t}{\sum_{t=1}^{T}h_t}
$
%\end{equation}
are typically employed, as they do not require any information in addition to the two time series $ {y}_{1:T} $ and $ {h}_{1:T} $ \cite{WHO2015}. However, they are well known to be affected by biases \cite{lipsitch2015potential}, particularly in the growing phase of the epidemic, when the number of hospitalisations increases more steeply \parencite{seaman2022estimating, seaman2022adjusting}. Corrections have been proposed, but they assume availability of further information, either about the time series of people recovered and discharged from hospital \cite{yip2005chain, yip1995some} or about the distribution of the delay between hospitalization and death or recovery \cite{ghani2005methods, overton2022novel}. 
This information is not likely to be available early in an outbreak, especially when a new pathogen emerges. For this reason it seems feasible to  rethink this problem in a \ac{SSM} context and integrate over the information that is not available.

{A \ac{SSM} can be formulated to} describe individuals staying and leaving the hospital over time. {Let $ X_t \in \Omega_{X_t}, Y_t\in \Omega_{Y_t} $ and $Z_t\in \Omega_{Z_t} $} denote the number of people that remain in hospital, die and recover (hence discharged) respectively during interval indexed by $ t $, with {$\Omega_{X_t}, \Omega_{Y_t}, \Omega_{Z_t}\subseteq\mathbb{N}$.} Moreover, let $ h_t $ denote the number of individuals entering the hospital at time $ t $; $ h_t $ is often known as it is monitored by the surveillance system. 

{Let $ \boldsymbol{\theta}=(p_\textsc{h},p_\textsc{d}, p_\textsc{r} ) $ be the static parameters, where $ p_\textsc{h},p_\textsc{d} $ and $p_\textsc{r}  $ are in the interval (0,1) and denote  }the probability of remaining in hospital, dying and recovering (and being discharged) within the interval considered (e.g. a week), respectively. Note that $  \boldsymbol{\theta}$ is subject to the constraint $ p_\textsc{h}+p_\textsc{d}+p_\textsc{r} =1 $.

The state process modelling the individuals remaining in hospital, dying and recovering can be expressed through the following chain Multinomial model:
\begin{equation}\label{eq:ssm}
(X_t, Y_t, Z_t| x_{t-1}, h_{t-1}) \sim \text{Multinomial} \left(x_{t-1}+h_{t-1}, \boldsymbol{\theta} \right) .
\end{equation}
{Hence,  throughout the time series the constraint:
\begin{equation*}
x_t+ y_t+ z_t=x_{t-1}+h_{t-1} \qquad \text{ for }t=1, 2, \dots, T
\end{equation*}
must always hold: i.e. the people that remained in hospital during interval $t-1$ ($x_{t-1}$) and those who entered during the interval $t-1$ ($h_{t-1}$) will either remain in hospital ($x_t$) or die ($y_t$) or recover and be discharged ($z_t$) during time interval $t$. }

As mentioned above, the number of individuals dying is very often closely monitored. Therefore, we assume that the deaths are fully observed %observation process is composed of a Dirac point mass distribution centred in the true number of individuals dying. 
and consider Equation \eqref{eq:ssm} as defining both the state and observation process. 
The state process at time $ t=0 $ depends on the disease under consideration and the monitoring system used. Since $ h_0 $, and every hospitalization before the start of the monitoring system, is not recorded, $ X_0 $ encapsulates all the individuals that were still in the hospital before the beginning of the data-collection. $ X_0 $, for the example of seasonal influenza, can be assumed to take non negative values, defined through, for example, the prior distribution:
\begin{equation}\label{eqx0}
X_0\sim \text{Poisson} (1.5).
\end{equation}

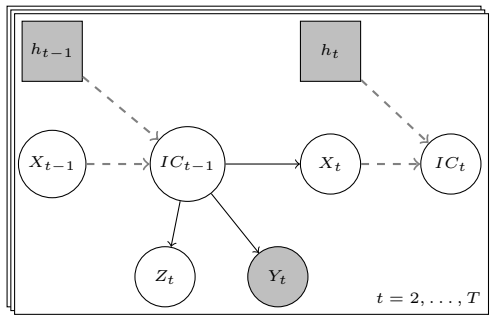
\begin{figure}[htp]
\centering
\vspace*{-0.3cm}   {\tiny
\centering
\begin{tikzpicture}[node distance=1.5cm]
\node (PL)[plate, xshift=-0.1cm , yshift=0.1cm]{};
\node (PL)[plate, xshift=-0.05cm , yshift=0.05cm]{};
\node (PL)[plate]{ };
\node (ICUtm1)[node, yshift=0cm, xshift=-.85cm] {$ IC_{t-1}$};
\node (ICUt)[node, right of=ICUtm1, xshift=2cm] {$ IC_{t}$};
\node (Yt)[nodeknown,below of=ICUtm1,yshift=0cm, xshift=1.2cm] {$ Y_{t}$};
\node (Xt)[node,left of=Yt] {$ Z_{t}$};
\node (Zt)[node,right of=Yt, yshift=1.5cm, xshift=-0.8cm] {$ X_{t}$};
\node (Ztm1)[node,left of=Xt, yshift=1.5cm] {$ X_{t-1}$};
%\node (Z0)[node,left of=Ztm1, xshift=-0.5cm,yshift=-1.5cm] {$ X_{0}$};
\node (Htm1)[compartment3,above of=Ztm1] {$ h_{t-1}$};
\node (Ht)[compartment3,above of=Zt] {$ h_{t}$};
\draw [->] (ICUtm1) -- node[anchor=south] { } (Xt);
\draw [->] (ICUtm1) -- node[anchor=south] { } (Yt);
\draw [->] (ICUtm1) -- node[anchor=south] { } (Zt);
%\draw [->] (Z0) -- node[anchor=south] { } (ICUtm1);
\draw [determ] (Ztm1) -- node[anchor=south] { } (ICUtm1);
\draw [determ] (Htm1) -- node[anchor=south] { } (ICUtm1);
\draw [determ] (Zt) -- node[anchor=south] { } (ICUt);
\draw [determ] (Ht) -- node[anchor=south] { } (ICUt);
\node (nc)[namecompartment,below of=ICUt, yshift=-0.3cm,  xshift=-0.3cm] {$t=2, \dots, T$};
\end{tikzpicture}}
\caption{\ac{DAG}  describing the \ac{SSM} of Equation \ref{eq:ssm}. Dashed arrows represent deterministic relationships and solid arrows stochastic relationships; white circles represent unknowns and grey boxes represent observed quantities, squares are assumed fixed while circles are assumed to distribute according to the observation process. }
\label{f1}
\end{figure}

The system is illustrated by the \ac{DAG} in Figure \ref{f1}, where a further node $ IC_t=h_t+x_t $ has been introduced to identify the size of the Multinomial \ac{r.v.}, i.e. the number of individuals in hospital at the start of the $ t $th time interval. 

While the time series $ {y}_{1:T} $ and $ {h}_{1:T} $ are known, we assume no observations on $ {x}_{0:T} $ ($ {z}_{1:T} $ follows from knowing $ {x}_{0:T} $). 
{In the remaining part of the paper we use \ac{SMC} methods to sample from the filtering distribution $ p({X}_{1:t}|{y}_{1:t}, \boldsymbol{\theta}) $ for $ t=1, \dots, T $, effectively marginalising over the unknown part of the process and providing an estimate of the likelihood $ \widehat{p}(\boldsymbol{\theta}| {y}_{1:T}) $. }

\section{The Lifebelt Particle Filter} \label{s3} 
{We here propose a \ac{SMC} algorithm aimed at avoiding particle collapse in discrete bounded \acp{SSM}. Particle collapse is avoided if, for all times $t=1,2,\dots, T$, there is at least one particle with positive weight $w^{(n)}_t$. This can be achieved by introducing one particle called the \textit{lifebelt particle}, with index $n=N$, for which Equation \eqref{eq:weights} in [Step 3] is:
\begin{equation}
w_{t}^{(N)}=\frac{p(y_t, x_t^{(N)}|x_{t-1}, \boldsymbol{\theta})}{q(x_t^{(N)}|x_{t-1})}>0 \qquad \text{ for } t=1, \dots T
\end{equation}
which becomes especially useful should all the other particles fail ($ w_{t}^{(n)}=0$ for all $n\neq N$).

Note that, for ease of notation, the explanation of the algorithm and the derivation of its exactness are outlined in the case of one lifebelt particle with index $n=N$, i.e. taking the last place in the ordered set of particles. However, the introduction of more than one lifebelt particle is discussed in Section \ref{ss4.3} and such particles could take any fixed predetermined location in the particle set.

This section proposes a method to include a lifebelt particle in a \ac{SMC} scheme, subsequently used to approximate the likelihood $p(y_{1:T}|\boldsymbol{\theta})$, which is summarised by:
\begin{itemize}
\item[(i)] assume the existence of a trajectory $x^{(N)}_{1:T}$ for which $p(y_t, x^{(N)}_{t}|x^{(N)}_{t-1})>0$ for all $t=1, \dots T$ and of an importance distribution $q_{\textsc{lb}}(x_t|x_{t-1})$ that, if iteratively applied, generates $x^{(N)}_{1:T}$;
\item[(ii)] at each time $t$, draw a set of particles and let one of them, the lifebelt, be proposed according to rule $q_{\textsc{lb}}$: $x^{(N)}_t \sim q_{\textsc{lb}}(\cdot|x^{(N)}_{t-1})$;
\item[(iii)] resample the particles so that the lifebelt particles is always retained, irrespectively of its weight $w^{(N)}_t$.
\end{itemize}

Point (i) is problem specific; as an example one could consider a bounded system such as a Binomial-Binomial model, given probabilities $ p_{\textsc{state}}$, $ p_{\textsc{obs}} \in (0;1)$, with state and observation distribution: 
\begin{equation*}
X_0= x_0, \qquad X_t \sim \text{Binomial}(x_{t-1}, p_{\textsc{state}}),\qquad Y_t \sim \text{Binomial}(x_{t}, p_{\textsc{obs}}) \qquad \text{ for }t=1, \dots,  T
\end{equation*}
where the two constraints that affect $x_t$ are $x_t\geq y_t$ and $x_t\leq x_{t-1} $; here a lifebelt trajectory could be formulated by letting $x_t$ taking as value the highest boundary of its space:  
\begin{equation*}
q_{\textsc{lb}}(x_t|x_{t-1})=\mathds{1}_{[x_{t-1}]}.
\end{equation*}

The remaining points are describes in Section \ref{s3.1} and \ref{s3.2},  respectively. 
}

\subsection{Deterministic mixture importance sampling and its sequential use}\label{s3.1}
{\Ac{DMIS} was firstly introduced by \cite{veach1995optimally} under the umbrella of \ac{MIS}, and later presented to the statistical community by  \cite{owen2000safe}}. In its static version, we are interested in a target distribution $ X $, whose \ac{p.d.f.} can be evaluated up to a normalising constant $ \pi(x) = \frac{f(x)}{\gamma} $ but from which we cannot directly sample. Samples from $ X$ can be obtained by importance sampling, where the importance distribution $ q(x) $ is defined as a deterministic mixture distribution. 
\begin{equation*}
\begin{split}
x^{(n)}&\sim q_1 \qquad \text{ for $ n=1, \dots{, } N_1 $}\\
x^{(n)}&\sim q_2 \qquad \text{ for $ n=N_1+1, \dots{, } N_1+N_2 $}\\
%		x^{(n)}_2&\sim q_2(x) \qquad \text{ for $ n=1, \dots{, } N_2 $}\\
& \dots\\
x^{(n)}&\sim q_G \qquad \text{ for $ n=\sum_{g=1}^{G-1} N_g+1, \dots{, } N $}
\end{split}
\end{equation*}
with $ N=\sum_{g=1}^{G} N_g$.
{The samples are pooled together and the weights are recomputed} as if the total sample was drawn from a mixture of all the densities $ q_1(x), q_2(x), \dots, q_G(x) $:
$ {q}(x^{(n)})= \frac{1}{N}\sum_{g=1}^{G}N_g q_g\left( x^{(n)}\right)  $
resulting in the deterministic mixture weight: 
\begin{equation*}\label{eqdmisw}
w^{(n)}= \frac{{f}(x^{(n)})}{\frac{1}{N}\sum_{g=1}^{G}N_g q_g(x^{(n)})} \qquad \text{for } n=1,2,\dots, N.
\end{equation*}
Despite the importance samples not being drawn from a mixture distribution, these weights allow exact \ac{MC} estimation (as derived in \cite{cornuet2012adaptive}). 

{\cite{cornuet2012adaptive} further expands \ac{DMIS} to allow for the temporal evolution of the weights and the components of the deterministic mixture. While this addition is not considered here, the paper importantly underlines that the samples obtained by \ac{DMIS}} are a valid importance-sampling approximation for the target $ \pi(x)$ if every sub-sample of size $ N_g $ is a valid importance sample, i.e. if the support of $ q_g $ contains the support of $ \pi(x)$. In this case, the \ac{DMIS} can be seen as a method that simply {pools} importance-sample estimators obtained from many different importance distributions. We extend this here by noting that if the support of the target density $\Omega $, can be covered {jointly} by the $ q_g $s:  i.e. $ \mathrm{supp}(q_g )=\Omega_g $, for $g=1, \dots, G  $, with $ \Omega\subseteq\bigcup_{g=1}^G \Omega_g $, then this importance sample retains exactness (see derivation in Appendix \ref{appA}).

{
Another valuable feature of \ac{DMIS} is that it is known to produce samples with lower variance compared with the ones obtained with standard importance sampling or using non-deterministic mixtures as importance distribution. This is proven theoretically and shown practically in \cite{elvira2019generalized}. 

Not only do these features make \ac{DMIS} a suitable candidate to draw importance samples in general, but also its stratified structure provides the perfect mechanism to draw the lifebelt particle together with another set of particles, with the security of always proposing a sample with positive weight. This can be obtained, for example, using the following deterministic mixture:
\begin{equation*}
\begin{split}
x^{(n)}&\sim q_1 \qquad  \qquad \text{    for } n=1, 2, \dots, N-1 \\
x^{(n)}&\sim q_2=q_{\textsc{lb}} \qquad \text{ for }  n=N
\end{split}
\end{equation*}
where $q_{lb}$ is a Dirac point mass distribution centred in the case-specific lifebelt particle. This would lead to weight:
\begin{equation}\label{eq:dmisw}
w^{(n)}= \frac{{f}(x^{(n)})}{\frac{N-1}{N} q_1(x^{(n)})+\frac{1}{N} q_2(x^{(n)})}  \qquad\text{ for }n=1,2, \dots, N.
\end{equation}
In the remainder of the paper the letter $q$ without indexes, which is used to denote a general importance distribution, identifies the chosen importance \ac{p.d.f.} and this include deterministic mixtures such as in  the denominator of Equation \eqref{eq:dmisw}.
}

\subsection{Combination of non-resampled and resampled particle sets}\label{s3.2}
A problematic task that arises when attempting to include the lifebelt particle in {a} particle filter is that, not only should the importance distribution at each time step be formulated as a \ac{DMIS}, but also the lifebelt particle trajectory should be {retained} over time, and therefore must be protected from disappearing in the resampling {step}. 
The weight presented in Equation \eqref{eq:weights} in [Step 1] needs to be further corrected to account for this merging of a set of particles drawn with \ac{SIRS} and another, the lifebelt, drawn with \ac{SIS} without resampling. 

{
More specifically, a new resampling step is formulated whereby the lifebelt particle $N$ undergoes two resampling procedures: firstly it is always resampled in the location $N$ of the particle set, and secondly, it can also be resampled in the remaining $N-1$ locations so that, should they all collapse, it would be able to repopulate the particle set. This scheme is implemented by artificially separating the lifebelt particle weight $w^{(N)}$ into two portions: one which is assigned to the location $N$ and the other which is used for resampling. A new parameter $r \in (0, 1)$ is, therefore introduced which determines the proportion of the lifebelt particle weight $w^{(N)}$ that is preserved from resampling and deterministically assigned to the lifebelt particle; this parameter is necessary for the sampler to retain unbiasedness (see proof in Appendix \ref{appB}). Any value of $r \in(0, 1)$ is valid, however higher values are to be preferred as they allow the lifebelt weight to remain as high as possible.} Assume that at time $ t-1 $ the set $ \left\lbrace {x}^{(n)}_{t-1}; \widetilde{w}^{(n)}_{t-1}\right\rbrace_{n=1}^N  $ is a {self-normalised} weighted sample from the filtering distribution. {[Step 1] of the \ac{SMC} routine, which produces the resampled set  $ \left\lbrace {x}^{(a_j)}_{t-1}\right\rbrace_{j=1}^N  $, consists of: first setting the $N$-th particle equal to the lifebelt particle:
\begin{equation*}
a_j=N \qquad\text{ for }j=N
\end{equation*}
and then sampling the remaining $N-1$ particles from all the set $\left\lbrace {x}^{(n)}_{t-1} \right\rbrace_{n=1}^N  $, (including the lifebelt particle) with resampling probabilities $p^{(n)}$ that accounts for the fact that some of the weight of the lifebelt particle was already allocated the deterministically-assigned particle in location $N$:

\begin{equation*}
a_j\sim \mathcal{C} \left\lbrace 1, 2, \dots , N ; p^{(1)}, p^{(2)}, \dots p^{(N)} \right\rbrace  \qquad\text{ for }j=1,2, \dots N-1
\end{equation*}
with \begin{equation}\label{eq:sampprob}
p^{(n)}=	\begin{cases}
\frac{\widetilde{w}_{t-1}^{(n)}}{1-\widetilde{w}_{t-1}^{(N)}r} & \qquad\text{for }n \neq N\\
&\\
{\widetilde{w}_{t-1}^{(N)}\cdot \frac{1-r}{1-\widetilde{w}_{t-1}^{(N)}r}} & \qquad\text{for }n = N\\
\end{cases}
\end{equation}
Note that weights $p^{(n)}$ are normalised by construction
\begin{equation*}
\begin{split}
\sum_{n=1}^{N} p_n &= \sum_{n=1}^{N-1}\frac{\widetilde{w}_{t-1}^{(n)}}{1-\widetilde{w}_{t-1}^{(N)}r} +	{\widetilde{w}_{t-1}^{(N)}\cdot \frac{1-r}{1-\widetilde{w}_{t-1}^{(N)}r}} \\
&= \frac{1}{1-\widetilde{w}_{t-1}^{(N)}r}\left[  \sum_{n=1}^{N-1}\widetilde{w}_{t-1}^{(n)}+	\widetilde{w}_{t-1}^{(N)}-r\widetilde{w}_{t-1}^{(N)}\right] = 1
\end{split}
\end{equation*}
since $\sum_{n=1}^N\widetilde{w}_{t-1}^{(n)}=	1$. 

Lastly, the un-normalised weights used for the likelihood estimation in Equation \eqref{eq:liket} account for this artificially unbalanced resampling scheme as follows:
\begin{equation}\label{eq:lbweights}
{w}^{(j)}_{t}= 
\begin{cases}
\frac{p\left( x_t,y_t|x_{t-1}^{(a_j)}\right) }{q\left( x_t|x_{t-1}^{(a_j)}\right) }\cdot\frac{1-\widetilde{w}_{t-1}^{(N)}r }{N-1} \cdot N &\qquad \text{for } j\neq  N\\
&\\
\frac{p\left( x_t,y_t|x_{t-1}^{(a_j)}\right) }{q\left( x_t|x_{t-1}^{(a_j)}\right)}\cdot	\widetilde{w}_{t-1}^{(N)}\cdot r\cdot N &\qquad \text{for } j=N.\\
\end{cases}
\end{equation}

Appendix \ref{appB} proves that the use of resampling scheme \eqref{eq:sampprob} and weights \eqref{eq:lbweights} leads to an unbiased estimation of the likelihood $\widehat{p}(y_{1:T}|\boldsymbol{\theta})$.
}

\subsection{Algorithm and features}\label{s3.3}
Algorithm \ref{alg1} reports the general form of the \ac{LBPF} with one lifebelt particle. This algorithm provides an unbiased estimate of the likelihood of the data $ y_{1:T} $ given a specific parameter value $\boldsymbol{\theta}$.

\begin{algorithm}[!h]
\SetKwInOut{Input}{input}\SetKwInOut{Output}{output}
\SetAlgoLined
\Input{$ {y}_{1:T} $; $ \boldsymbol{\theta} $; $ N $; $r$; $ q_1 $; $ q_2 $; $ p(x_0) $, $x_{0}^{*}$.}
\Output{an approximation of the likelihood $ \widehat{p}({y}_{1:T}|\boldsymbol{\theta}) $.}
\BlankLine
\For{$n=1, \dots,N-1$}{	
% resample\\
$ {x}_{0}^{(n)}\sim p(x_0)$ \hfill sample the state at time 0\\}
\If{$n=N$}{	
% resample\\
$ {x}_{0}^{(n)}=x_{0}^{*}$ \hfill set the state at time 0 to value $x_{0}^{*}$ \\}
$ \widetilde{w}^{(n)}_0 = \frac{1}{N} $ \hfill initialise the weights\\
\For{$t = 1, \dots, T$}{
\For{$n=1, \dots,N-1$}{	
% resample\\
$ 	a_n\sim \mathcal{C} \left\lbrace 1, 2, \dots , N ; p^{(1)}, p^{(2)}, \dots p^{(N)} \right\rbrace $ \hfill resample w.p. \eqref{eq:sampprob}\\
%propagate:\\
$ {x}_{t}^{(n)}%| {x}_{t-1}^{(n)}, y_t
\sim q_1 \left(x_t|{x}_{t-1}^{(a_n)} , y_t\right) $\hfill propagate\\
%compute the weights:\\
$ {w}_{t}^{(n)}=\frac{p\left( x_t,y_t|x_{t-1}^{(a_n)}\right) }{q\left( x_t|x_{t-1}^{(a_n)}\right) }\cdot\frac{1-\widetilde{w}_{t-1}^{(N)}r }{N-1} \cdot N $ \hfill compute weights \eqref{eq:lbweights} \\}
\If{$ n=N$}{
$ a_n=N $ \hfill {set the {lifebelt} particle deterministically}\\
$ {x}_{t}^{(n)}	\sim q_2 \left(x_t|{x}_{t-1}^{(a_n)}, y_t\right) $ \hfill propagate \\
$ {w}_{t}^{(n)}=\frac{p\left( x_t,y_t|x_{t-1}^{(a_n)}\right) }{q\left( x_t|x_{t-1}^{(a_n)}\right)}\cdot	\widetilde{w}_{t-1}^{(N)} \cdot r \cdot N
$ \hfill compute the weight \eqref{eq:lbweights}}

$\widetilde{w}_t^{(n)}=\frac{{w}_t^{(n)}}{\sum_{n=1}^N {w}_t^{(n)}} \qquad  \text{for } n=1,\dots, N$\hfill	normalize the weights}
Approximate the likelihood by 
$  \widehat{p}(y_{1:T}|\boldsymbol{\theta})\approx \prod_{t=1}^{T} \frac{1}{N} \sum_{n=1}^{N}{w}^{(n)}_t
$
\caption{Lifebelt Particle filter}
\label{alg1}
\end{algorithm}

The \ac{LBPF} aims at being more robust than traditional \ac{SIRS} and \ac{SIS} schemes, thanks to the joint effect of these two safety measures: firstly the proposal of one lifebelt particle is guaranteed thanks to the use of \ac{DMIS}; secondly, the proposed resampling scheme guarantees the preservation of the lifebelt particle for all the steps from $t=1$ to $t=T$.

The choice of components for the deterministic mixture proposal allows for great flexibility: ideally one would choose $q_1$ so that it would perform well in high-density regions of $\boldsymbol{\theta}$ and a minimal amount of resources are spent on $q_2$ which, by construction, is likely to be anomalous (e.g. taking the boundary of the space of $x_{1:T}$). Should all the particles sampled with $q_1$ collapse, the lifebelt particle would come to the rescue and allow the \ac{SMC} routine to continue until the end of the time series.

\section{Application of the LBPF to the hospitalisation-death model} \label{s4}
The estimation problem introduced in Section \ref{ss2.3} can be initially tackled by  formulating a \ac{SIRS} that uses an importance distribution $q$ which accounts for the data at time $y_t$. As this candidate particle filter fails, an  \ac{LBPF} is adopted and described (Section \ref{ss4.2}), followed by  a further extension, that allows for the presence of multiple lifebelt particles (Section \ref{ss4.3}).

\subsection{A failing particle filter for $ \boldsymbol{\theta} $}
A \ac{SIRS} is used to sample from the unknown number of people staying in the hospital $x_{t}$ for $t=0, \dots T$. In particular, an \ac{AuPF} can be formulated using $ y_t $, the death observed at time $ t $, to inform the importance distribution $q(x_t|x_{t-1})$. Given system (\ref{eq:ssm}), conditionally on the number of people dying at time $ t $, $ y_t $, and on the people currently in hospital $ x_{t-1}+h_{t-1} $, the number of people remaining in hospital $ x_t $, follows a Binomial \ac{r.v.} with size $ x_{t-1}+h_{t-1} +y_t$ and probability $ p_\textsc{h}/(1-p_\textsc{d}) $. Therefore, a suitable importance distribution  for $x_t$ is:
\begin{equation}\label{eq:q1}
q(x_t|y_t,x_{t-1}, \boldsymbol{\theta}) \sim \text{Binomial} \left(x_{t-1}+h_{t-1} +y_t;   \frac{p_\textsc{h}}{1-p_\textsc{d}} \right).
\end{equation}
At each time step, a particle set of size $ N $ is proposed according to \eqref{eq:q1}, each particle is then assigned un-normalised weight \eqref{eq:weights},  with $ p(\cdot) $ being the state-observation \ac{p.d.f.} \eqref{eq:ssm} and $ q(\cdot) $ the importance \ac{p.d.f.} \eqref{eq:q1}; these weights contribute to the approximation of the likelihood of the parameter $ \boldsymbol{\theta} $ as per Equation \eqref{eq:like1T}.

\begin{figure}[h]
\centering
\begin{subfigure}[b]{0.46\textwidth}
\centering
\hspace*{-2.5cm}\includegraphics[scale=.8]{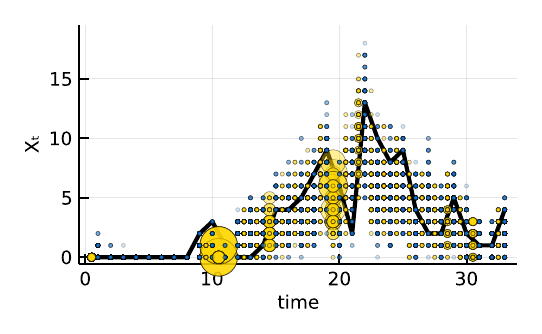}
\caption{ }
\label{fig2a}
\end{subfigure}\begin{subfigure}[b]{0.46\textwidth}
\centering
\vspace*{-0.3cm}\includegraphics[scale=.8]{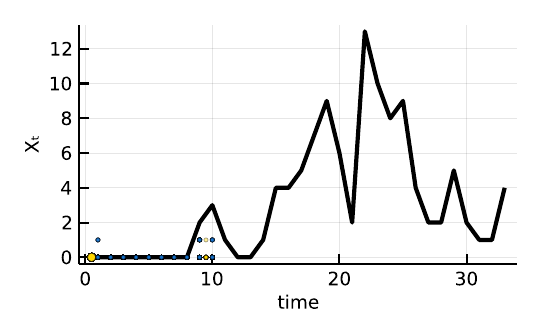}
\caption{ }
\label{fig2b}
\end{subfigure}
\caption{Particle swarm over time (x-axis) of one application of a \ac{SIRS} algorithm with importance distribution \eqref{eq:q1} to data generated with parameters $\boldsymbol{\theta}=(0.1, 0.8, 0.1)  $, with $ N=500 $, for estimating the likelihood of (a) the parameters generating the data and (b) a value in the tail: $\boldsymbol{\theta}=(0.01, 0.6, 0.39)  $. The black line in the background is the true, unknown, value of the latent state $ {x_{1:T}} $; the yellow dots are the samples drawn from the importance distributions at each time step with their size proportional to the normalised weight; the equally-sized blue dots are the resampled particles.}
\label{fig2}
\end{figure}

Figure \ref{fig2} illustrates the performance of this \ac{SMC} routine when applied to simulated data (where the true latent process $x_{0:T}$ is known).  The particle-swarm is directed to target the filtering distribution, accounting for the value of the current observation, $ y_t $. However, the chained nature of the system is such that the same particles that should be the most compatible with the observation $ y_t $ might also be incompatible with future outcomes $ y_{t+1}, y_{t+2}, \dots $. 
Since the distributions are discrete and on constrained spaces, this means that the size parameter of the Multinomial can be lower than a realisation (i.e., the observation). This incompatibility is likely to happen, for example, when estimating the likelihood of a parameter value $ \boldsymbol{\theta} $, that is far from the true value. Figure \ref{fig2} shows the particles proposed and resampled only with importance distribution \eqref{eq:q1}. Particle collapse is caused by the fact that, for all 500 particles, the number of people remaining in hospital at time $ t=14 $, $ X_{14} $, is smaller than the observation at that time. 

An algorithm such as the \ac{APF} would only try and overcome this particle-collapse problem via an ever-increasing number of computations (i.e. reusing the same importance distribution an increasing number of times), without really addressing the issue of an importance distribution that, while wisely chosen, is not robust enough to allow for outliers. This motivates the creation of a mechanism to sample from the filtering distribution that goes beyond the intensive use of the importance distribution \eqref{eq:q1}. 

\subsection{The lifebelt particle in action}\label{ss4.2}

In the example presented in Section \ref{s2}, there is a clear definition of the boundary of the state process. The upper limit of the state process is determined by the situation when no one leaves the hospital except for the observed individuals that die, ($ {y}_{1:T} $). If this is the case the state process $ {X}_{1:T} $ takes its highest possible value at each time step. 

A rule can be set to obtain a realisation of this specific trajectory that delimits the boundary of the state-space:
\begin{equation}\label{eq:lbp}
\begin{cases}
{x}_{t}&={x}_{t-1}+h_{t-1}-y_t\\
y_t&=y_t\\
z_t&=0
\end{cases}
\end{equation}

This trajectory has the special characteristic that it cannot collapse to zero weight, since, not only does it guarantee that the number $ x_t $ remains higher than the current observation $ y_t $, but it takes its highest possible value, preventing future collapses due to the constrained space of the Multinomial. In this sense, a lifebelt particle that satisfies assumption (i) of Section \ref{s3} can be designed according to rule \eqref{eq:lbp}. 

We consider the combination of $ N-1 $ particles drawn according to the  importance distribution \eqref{eq:q1} and $1 $ lifebelt particle. The \ac{DMIS} has then two components:
\begin{enumerate}
\item $ x_t^{(n)}\sim q_1(x_t^{(n)}) $ is the Binomial distribution that is obtained conditionally on the number of deaths at time $ t $, $ y_t $, being available as per Equation \eqref{eq:q1}
for $ n=1, \dots, N-1 $;
\item $ x_t^{(n)}\sim q_2(x_t^{(n)}) $ is a Dirac point mass distribution centred in the {lifebelt} case where all the people who enter the hospital and do not die stay in the hospital:
\begin{equation}\label{eq:q2}
{x}_{t}^{(n)}| {x}_{t-1}^{(n)}, y_t \sim \delta_{{x}_{t-1}^{(n)}+h_{t-1}-y_t} \left(x\right)
\end{equation} 
for $ n=N$.
\end{enumerate}

\begin{figure}[h]
\centering
\centering
\begin{subfigure}[b]{0.46\textwidth}
\centering
\hspace*{-2.5cm}\includegraphics[scale=.8]{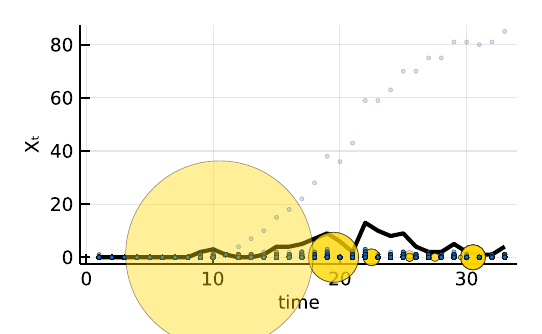}
\caption{ }
\label{fig3a}
\end{subfigure}\begin{subfigure}[b]{0.46\textwidth}
\centering
\vspace*{-0.3cm}\includegraphics[scale=.8]{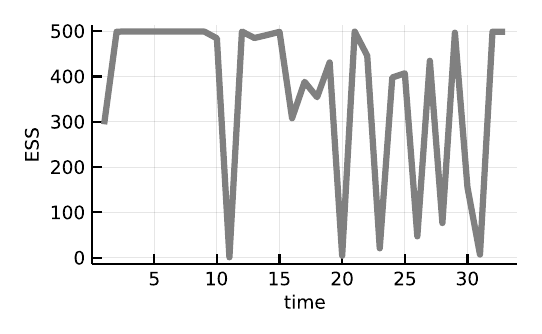}
\caption{ }
\label{fig3b}
\end{subfigure}
\caption{(a) Particle swarm of one application of Algorithm \ref{alg1} with $ N=500 $, for estimating the likelihood of the parameter $\boldsymbol{\theta}=(0.01, 0.6, 0.39)  $, applied to data generated with parameter $\boldsymbol{\theta}=(0.1, 0.8, 0.1)  $. The black line in the background is the true, unknown, value of the latent state $ \boldsymbol{x_{1:T}} $; the yellow dots are the samples drawn from the importance distributions at each time step, their size is proportional to the normalised weight; the equally-sized blue dots are the resampled particles. (b) ESS of the particles over time. }
\label{flbpf}
\end{figure}

The importance distribution $ q_2 $ generates the {lifebelt} particle and ensures it will always have a positive weight, hence it will not be resampled but always progressed. An example of the proposed and accepted particles are displayed in Figure \ref{flbpf}.

\subsection{The lifebelt fleet: a richer LBPF}\label{ss4.3}
\begin{figure}[h]
\centering
\includegraphics[scale=1]{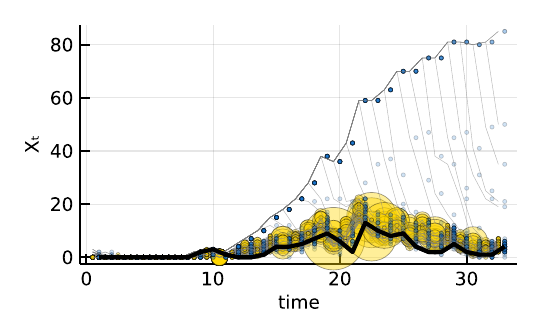}
\caption{Particle swarm obtained by the use of $T+1$ lifebelt particles, one of which follows deterministic dynamics given by \eqref{eq:q2}, and the remaining $ T $ initially follow the dynamics \eqref{eq:q1}, and as time progresses rejoin the main particle swarm. The thick black line is the true, unknown, value of the latent state $ \boldsymbol{x_{1:T}} $; the yellow dots are the samples drawn from the importance distributions at each time step, their size is proportional to the normalised weight; the equally-sized blue dots are the resampled particles. The trajectories are connected by grey lines according to the particle indices. }
\label{ffleet}
\end{figure}

Note that this algorithm is flexible to many variations. We have presented a specific \ac{LBPF} where only one particle follows non-resampling dynamics. Another proposal could be to have more than one lifebelt particle and allocate them so that all the upper tail of the hidden process is covered. For example one could start  with $ T+1 $ particles in the lifebelt location, where $ T $ is the length of the time series modelled. At each time point one of the $ T+1 $ particles that were in the lifebelt location starts following the dynamics of the main particle swarm. These particles can be visualised in Figure \ref{ffleet}.

This would mean introducing a further importance density to the \ac{DMIS} proposed in Section \ref{s3.1}:

\begin{enumerate}
\item[3.] $ x_t^{(n)}\sim q_3(x_t^{(n)}) $ depends on time and it is different for each particle: as the particle index $ n $ increases, the particle will stop following $ q_2 $ and follow $ q_1 $ instead. \\
Let $ k=n-N+T+2 $

\begin{equation*}\begin{cases}
{x}_{t}^{(n)}| {x}_{t-1}^{(n)}, y_t \sim \delta_{{x}_{t-1}^{(n)}+h_{t-1}-y_t} \left(x\right)	& \text{if }k<t\\
{x}_{t}^{(n)}| {x}_{t-1}^{(n)}, y_t \sim \text{Binomial} \left( ({x}_{t-1}^{(n)}+h_{t-1}-y_t), \left( 1-\frac{p_\textsc{r}}{1-p_\textsc{d}}\right) \right)	& \text{if }k\geq t\\
\end{cases}
\end{equation*} 
where $ n=N-T-1, N-T, N-T+1, \dots, N-1 $. 
\end{enumerate}

The choices of how to formulate \ac{DMIS} are problem-dependent and should be tailored to the hidden space under analysis. The advantage of a particle filter constructed under these rules is that the amount of particles employed for some particular tasks (e.g. sampling from the boundaries or the tails of the latent space) are specified. Differently from algorithms like the \ac{APF}, the computations have always an upper bound and the statistician is free to decide where to target their computational resources. 

The reason for choosing multiple lifebelt particles is also the fact that, in case one of the lifebelt particles is resampled, if all other particles fail, its weight might be very small. This is due both to the fact that (a) the lifebelt particle often represents an extreme scenario, and (b) its weight is decreasing quickly as it does not go through resampling. Multiple lifebelt particles might allow us to cover the latent space better and retain less extreme particle trajectories. Note that, even with the use of the \ac{LBPF}, as happens in many \ac{SMC} applications, there could be a numerical-precision error whereby small weights are approximated to 0 and the approximation of the log likelihood can be $ -\infty $, in which case this means that effectively the parameter of which we are evaluating the likelihood is not supported by the data. The \ac{LBPF} allows a better approximation of the tails of the likelihood than other particle filtering methods. 

\subsection{Use of the LBPF in a pseudo marginal method}\label{s4.2}

The \ac{LBPF} provides an unbiased estimate of the likelihood of the data $ {y}_{1:T} $ given a specific value of the parameter vector $ \boldsymbol{\theta} $ (see Appendix \ref{appB}).
For this reason it can be used within a pseudo-marginal method \cite{andrieu2009pseudo}. Specifically, the posterior distribution $p(\boldsymbol{\theta}|y_{1:T})$ is sampled according to a \ac{PMMH} algorithm as illustrated in \cite{andrieu2010particle}. % as it retains exactness even with a finite number of particles. 

To explore $ \boldsymbol{\theta}=(p_\textsc{h}, p_\textsc{d}, p_\textsc{r}) $, the parameter space was transformed to lie in $ \mathbb{R}^2 $ 
\begin{equation*}\label{key}
\begin{split}
\gamma_1 &= \text{logit}\left( \frac{p_\textsc{d}}{p_\textsc{d}+p_\textsc{r}}\right) \\
\gamma_2&= \text{logit}(p_\textsc{d}+p_\textsc{r})
\end{split}
\end{equation*}
with inverse:
\begin{equation*}
\left( \frac{e^{\gamma_1+\gamma_2}}{(1+e^{\gamma_1})(1+e^{\gamma_2})}, \frac{e^{\gamma_2}}{(1+e^{\gamma_1})(1+e^{\gamma_2})}\right) .
\end{equation*}
A Dirichlet(1,1,1) was assumed on the simplex of $ \boldsymbol{\theta} $, granting a uniform prior. 
A dataset was generated with a specific parameter set $\boldsymbol{\theta}=( p_\textsc{h}=0.3, p_\textsc{d}=0.5, p_\textsc{r}=0.2 )$.  An \ac{MCMC} was run: convergence is fast and the chains show good mixing (see Figure \ref{fig5b} below). 
\begin{figure}[!h]
\centering
\includegraphics[scale=.7]{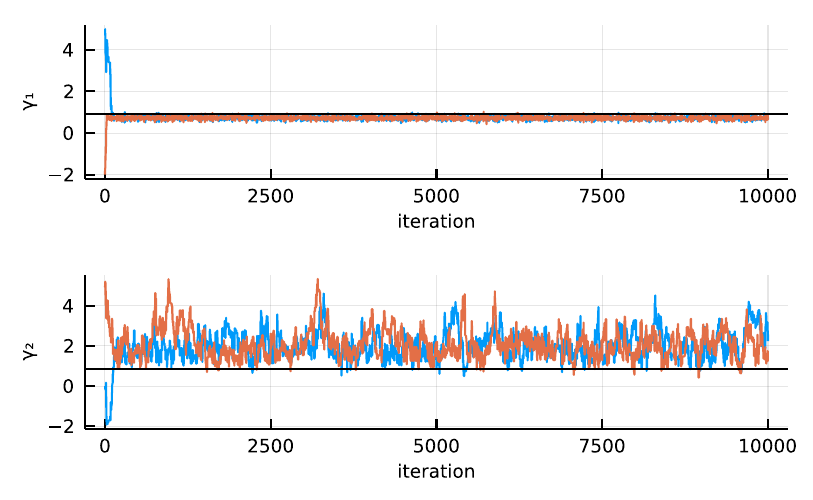}
\caption{pMCMC with likelihood estimated by Algorithm \ref{alg1}. Two parallel chains starting from separate  initial values are displayed with two colours.}
\label{fig5b}
%	\end{subfigure}
%	\caption{}
%	\label{f5}
\end{figure}

Compared to the standard pMCMC that adopts a \ac{SIRS} using only \eqref{eq:q1} for the estimation of the likelihood, there is no apparent difference in the chain and, as expected, no bias is introduced in the posterior inference (in accordance with the proof reported in Appendix \ref{appB}). 

\section{Comparison with Alive Particle Filter}\label{s5}
In this section we assess whether the \ac{LBPF} presented in Section \ref{ss4.2} with $N=500$ total particles does indeed cut the computation time compared to the \ac{APF}. 

The \ac{APF} allows the number of proposed particles at each time step, $n_t$, to increase until a desired number of particles which are ``good matches" is obtained. The number of particles $n_t$ is distributed as a Negative Binomial \ac{r.v.} as it identifies the number of attempts needed to obtain $N+1$ successes. 
Note that the \ac{APF} is a good candidate filter for any problem where a ``good match"  can be defined, whereby the success of a particle can be discretely identified. This makes it particularly suitable for \ac{ABC}, where a particle is successful if synthetic observations lie within a distance $\varepsilon$ of the true observations, and for discrete bounded problems, such as ours, where the observation probability is discrete and take value either 0 or the few positive values of the observation \ac{p.d.f.} (see Remarks 1-4 in \cite{drovandi2016exact}). 

We implemented a version of the \ac{APF} that uses as importance distribution Equation \eqref{eq:q2}. Moreover we consider a target number of non-zero-weighted particles of $ N=500 $, and we bound the number of proposals at each time point to be $ n_t\leq 1 $ million, i.e., when 500 non-zero-weighted particles are not obtained by the 1 millionth attempt, the filter proceeds to the next time step resampling only among the simulated particles, if any of these has a weight larger than 0. 

We analyse a simulated dataset, generated with parameter set: $\boldsymbol{\theta}=( p_\textsc{h}=0.3, p_\textsc{d}=0.5, p_\textsc{r}=0.2 )$. This dataset presents some challenges as there are many spikes in the unknown number of people remaining in hospital, which might be difficult to be matched by the importance density. 

\begin{figure}[h]
\centering
\includegraphics[scale=.6]{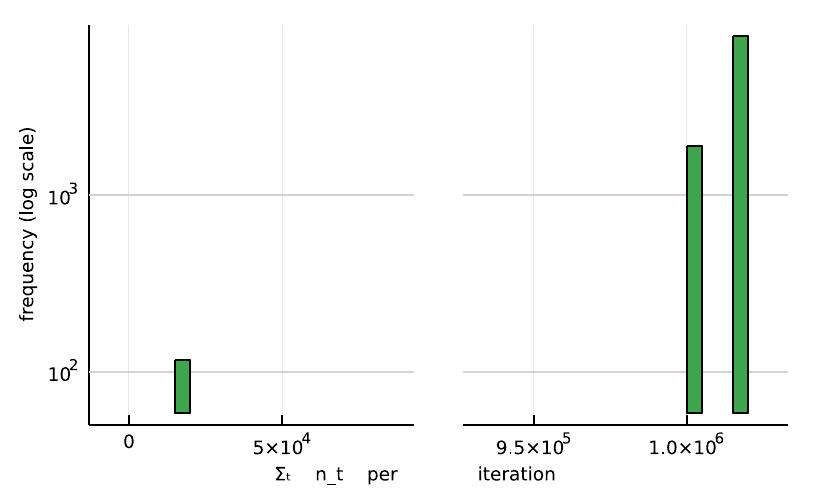}
\caption{Histogram of the number of attempts made by the \ac{APF}.}
\label{fig6}
\end{figure}

While the chain in practice looks very similar to the one obtained with the \ac{LBPF}, the \ac{MCMC} using \ac{APF} takes 30\% to 50\% more time than the \ac{MCMC} that uses the \ac{LBPF}. To understand why this is happening in practice we can look at the distribution of the total number of proposals over time $ n=\sum_{t=1}^{T}n_t $ for the proposed parameter $ \boldsymbol{\theta}^* $  at each iteration of the \ac{MCMC}; the histogram of this quantity over the 10,000 \ac{MCMC} iterations is reported in Figure \ref{fig6}. 

The small column to the left-hand side represents those \ac{MCMC} iterations for which the \ac{APF} obtains an estimate of the likelihood only using 500 particles per time point. 
The second column from the right, in the middle represents all those \ac{MCMC} iterations for which the \ac{APF} failed in its attempt, at some point $ t<T $, to obtain any non-zero-weighted particles. The filter stopped at $ t $ and estimated the likelihood of the parameter proposed in that \ac{MCMC} iteration to be zero. Lastly the highest column on the right-hand side represents all those \ac{MCMC} iterations for which the \ac{APF} managed to obtain at least one particle with non-zero weight, it continued to propose until the maximum limit of 1 million particles was reached, and then resampled those few non-zero-weighted particles to continue the filter until the end of the time series $ T $. 

If we compare this with the \ac{LBPF}, we note that the latter always uses a constant number of particles (in this case fixed to 500). When the particles from the main filter do not manage to provide a good approximation of the filtering distribution, instead of insisting with the intensive use of an inadequate importance distribution, it samples the lifebelt particle, that is formulated to be more robust against particle failure. 

\begin{figure}[h]
\centering
\hspace*{-1.0cm}\includegraphics[scale=0.55]{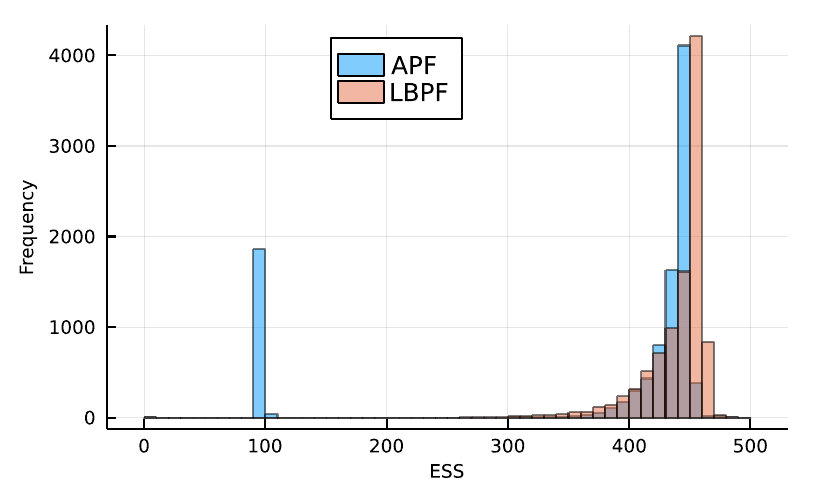}\includegraphics[scale=0.55]{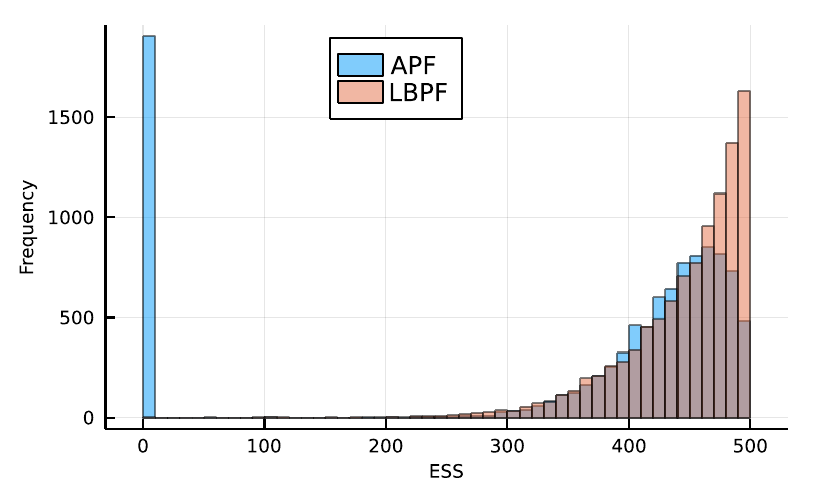}
\caption{Comparison of the \ac{ESS} of the particles averaged over time (left) and at the last time step (right) for the parameter proposed at each iteration of the MCMC run with \ac{APF} and \ac{LBPF}.}
\label{fig7}
\end{figure}

Irrespective of the computation time, we could compare the quality of the likelihood approximation via \ac{LBPF} vs via \ac{APF} for each proposed value of the \ac{MCMC}. This quality can be assessed by computing the \ac{ESS} of the particles used to approximate the likelihood. Figure \ref{fig7} compares the \ac{ESS} of the particle swarm used to approximate the value of the proposed value $\boldsymbol{\theta}^*  $ at each iteration of the \ac{MCMC}. Both considering the \ac{ESS} averaged across time points and the \ac{ESS} of the particles at the final time point, \ac{LBPF} seems to provide better estimates of the likelihood in terms of higher \ac{ESS}. Unlike the \ac{APF}, the \ac{LBPF} does not have a mode at small \ac{ESS} values where the particle filter collapses and the likelihood is estimated to be zero. Furthermore, it seems that when the lifebelt particle is actually used, this regenerates and enriches the particle swarm in the following time steps as noted in Section \ref{s4}. 

Note that these results refer to the specific scenario considered, which happens to be particularly challenging, rather than providing a general conclusions on the performance of the \ac{LBPF} versus the \ac{APF}. Moreover, we have used the basic formulation of the \ac{APF}, without considering more complex/bespoke versions of this algorithm. 

\section{Discussion}\label{s6}

In this paper we have proposed a new particle filter that combines \ac{SIRS} with \ac{SIS} without resampling whose importance distribution is chosen specifically to avoid particle failure. The algorithm relies on the introduction of a specific {lifebelt} particle, robust to particle collapse, whose formulation is problem specific.

The use of deterministic mixtures in the importance distribution allows the particle filter to always propose the  value of the lifebelt particle at time $t$ and our novel resampling scheme allows the lifebelt particle to survive until the end of the time series $T$. While previous work has exploited deterministic mixtures to formulate robust particle filters \cite{klaas2005toward, kronander2014robust, elvira2018search, branchini2021optimized}, to our knowledge this is the first time that they are combined to a resampling scheme that effectively creates a combination of resampled and non-resampled particles.

We have presented the example of a \ac{SSM} for hospital admissions and deaths in hospital where small counts and strong temporal dependence makes running a standard particle filter more challenging. 

There are many other instances in which the \ac{LBPF} could be used. For example, one could consider queue systems where entries, exits and waiting times are only partially monitored, here the lifebelt particle could be formulated according to a criteria similar to the one used for the deaths in hospital (i.e. covering the upper limit of the hidden state). Other examples could be taken from the inference of epidemic models \cite[as in][]{mckinley2020efficient}: here the individuals are divided into groups according to their disease status and observations are often made on the individuals leaving the infectious state. The lifebelt particle could be formulated by assuming that all the individuals who are eventually ill, become infectious soon after the epidemic has started and progressively move to the recovered/removed state. Lastly, the \ac{LBPF} could also be extended to continuous \acp{SSM}. In this context the lifebelt particle could follow an over-dispersed importance distribution to protect against future shocks. While this does not guarantee absolute safety of the lifebelt particle, it may still be effective at preventing particle collapse in many cases. The extension to continuous \ac{SSM}, however, would require a further analysis of exactness, since the proofs presented here are based on discrete \acp{r.v.}. 

Our contribution, rather than providing a tool that can be blindly applied to any \acp{SSM}, offers a framework that can be used to combine different importance distributions, some of which might not be formulated to sample from the mass of the filtering distribution, but instead have the role of covering its tail. This framework becomes useful under a computational budget perspective: the resources are better shared between multiple candidate importance distributions. Moreover the proposed algorithm presents advantages also under the point of view of inferring the hidden space: preserving particles that sample from the tails of the filtering distribution leads to a better estimate of the likelihood in terms of effective sample size. We have shown how the combination of multiple importance distributions is possible without biasing the likelihood estimator. 

In addition to proposing this setup which helps in many other settings, the example presented of the estimation of \ac{CFR} for hospitalised patients is of key relevance. At the beginning of the recent Covid-19 pandemic, only time series of cases and deaths were reported. Ratios of cumulative counts of cases and deaths provided biased estimates of the \ac{CFR}: our particle filter approach instead integrates over what is unknown and, given the introduction of the lifebelt particle, is robust to data which might be unexpected. 

\section*{Acknowledgements}
AC, GOR, SEFS and DDA are supported by Bayes4Health EPSRC Grant EP/R018561/1. AC, AMP, PJB and DDA are supported by the MRC, programme grant MC\_UU\_00002/11. TJM is supported by an ``Expanding Excellence in England" award from Research England.. AC is further supported by by the Royal Society (Dimension Supercharged Projective Sampling project). {We are thankful to the two unknown reviewers that provided insightful comments to a previous version of this manuscript. }

\appendix
\section{Exactness of DMIS}\label{appA}
\begin{theorem} \label{t1}
Given a sample $\left\lbrace x^{(n)}\right\rbrace _{n=1}^{N}$ drawn according to \acl{DMIS} ( \ac{DMIS}) with importance distributions $q_1, \dots, q_G$ with supports $\Omega_g$ for $g=1, \dots G$ and a function $h(X)$ of the random variable $X$, with $
\mathbb{E}\left[ h(X)\right]=\mu $; then the estimator:
\begin{equation*}
\hat{I} = \frac{1}{N} \sum_{g=1}^{G} \sum_{n=1}^{N_g} h\left(x^{(n)}\right)\frac{{f}(x^{\left(n\right)})}{\frac{1}{N}\sum_{l=1}^{G}N_l q_l\left(x^{(n)}\right)} .
\end{equation*} 
is unbiased, i.e.:
\begin{equation*}
\mathbb{E}\left[ \hat{I}\right]=\mu .
\end{equation*}
\end{theorem}

\begin{proof}
The expected value of the \ac{DMIS} estimator is:
\begin{equation*}
\begin{split}
\mathbb{E}\left[\hat{I} \right]&= \mathbb{E} \left[\frac{1}{N} \sum_{g=1}^{G} \sum_{n=1}^{N_g} h\left(x^{(n)}\right)\frac{{f}(x^{(n)})}{\frac{1}{N}\sum_{l=1}^{G}N_l q_l\left(x^{(n)}\right)} \right]\\
&=\frac{1}{N} \sum_{g=1}^{G} \sum_{n=1}^{N_g} \mathbb{E} \left[ h\left(x^{(n)}\right)\frac{{f}(x^{(n)})}{\frac{1}{N}\sum_{l=1}^{G}N_l q_l\left(x^{(n)}\right)} \right]
\end{split}
\end{equation*}
due to linearity of $\mathbb{E}$.\\
The $ N_g $ samples are \ac{iid} draws from the $ g $-th importance distribution. \ac{DMIS} assumes that each $ q_g $ is an appropriate importance distribution for $ f $, i.e. $ \Omega \subseteq \Omega_g$, the domain of $ f $. Exactness, however, holds also if this is not the case.
\begin{equation*}\begin{split}
\mathbb{E}\left[\hat{I} \right]&=\frac{1}{N} \sum_{g=1}^{G} {N_g} \sum_{x\in \Omega_g} h\left(x\right)\frac{{f}(x)}{\frac{1}{N}\sum_{l=1}^{G}N_l q_l\left(x\right)}q_g\left(x\right)\\
&\\			\end{split}
\end{equation*}
{Note that }$f(x)=0${ outside its support, i.e. in the set }$ \Omega_g\setminus\Omega$ { and that }$q_g(x)=0${ outside } $\Omega_g$. Hence, we can add marginalizing sums that are equal to 0 to cover all the support $ \Omega$.
\begin{equation*}\begin{split}
\mathbb{E}\left[\hat{I} \right]&=\sum_{g=1}^{G} {N_g} \left[\sum_{x \in\Omega \cap \Omega_g} h\left(x\right)\frac{{f}(x)}{\sum_{l=1}^{G}N_l q_l\left(x\right)}q_g\left(x\right) + \underbrace{\sum_{x\in\Omega\setminus\Omega_g} h\left(x\right)\frac{{f}(x)}{\sum_{l=1}^{G}N_l q_l\left(x\right)}q_g\left(x\right) }_{=0} + \right. \\
& \qquad \qquad \qquad \qquad \left. + \underbrace{\sum_{x\in\Omega_g\setminus\Omega} h\left(x\right)\frac{{f}(x)}{\sum_{l=1}^{G}N_l q_l\left(x\right)}q_g\left(x\right)}_{=0}\right]\\
&=\sum_{g=1}^{G} {N_g} \sum_{x\in\Omega} h\left(x\right)\frac{{f}(x)}{\sum_{l=1}^{G}N_l q_l\left(x\right)}q_g\left(x\right) \\			\end{split}
\end{equation*}
{Given that the sums are all on the same support } $\Omega${ we can bring the outer summation inside }\\
\begin{equation*}\begin{split}
\mathbb{E}\left[\hat{I} \right]&= \sum_{x\in\Omega} h\left(x\right){f}(x)\sum_{g=1}^{G} {N_g}\frac{q_g\left(x\right)}{\sum_{l=1}^{G}N_l q_l\left(x\right)} \text{d}x\\
&= \sum_{x\in\Omega} h\left(x\right){f}(x) =\mu\\
\end{split}
\end{equation*}
and exactness of the estimator is proved. 
\end{proof}

Note that, while \cite{cornuet2012adaptive} assumed that $ \Omega \subseteq \Omega_g$ was a condition for exactness, this can be extended to the case where the $ G$ $ q_g $s jointly cover the support of the target density, i.e. $ \Omega=\bigcup_{g=1}^G \Omega_g $, while not containing it. In  fact all the steps above are valid also while assuming $ f(x)>0 $ in $ \Omega\setminus\Omega_g $ and $ q_g(x)=0 $ in $ \Omega\setminus\Omega_g $.

\section{Unbiasedness  of \ac{SMC} weights}\label{appB}

This Appendix contains the proof that the joint use of sampling probabilities \eqref{eq:sampprob} and un-normalised weights \eqref{eq:lbweights} leads to an unbiased estimator of the likelihood $p(y_{1:T}|\boldsymbol{\theta})$ in a similar fashion to  \cite{pitt2012some}. As all the proof is valid given a specific value of $\boldsymbol{\theta}$, this parameter is omitted throughout and consider constant and given (as it is the case, for example, in the step of a \ac{PMMH}).

Let $y_{1:T}$ and $x_{0:T}$ be discrete \acp{r.v.} and denote with $\Omega_t$ the support of $X_t$ for all $t=0, 1, \dots, T$

\begin{theorem}\label{theo:unbiased}
	Let $\mathcal{A}_t = \left\lbrace x_t^{(n)}, {w}_t^{(n)}\right\rbrace_{n=1}^{N} $ be the swarm of particles at time $t$, so that, with normalised weights $\widetilde{w}_t^{(n)}=\frac{{w}_t^{(n)}}{\sum_{l}^{N}{w}_t^{(l)}}$, it provides a Dirac-delta approximation of the filtering distribution at time $t$. Then:
	\begin{equation}
		\widehat{p}(y_{1:T})= 	\widehat{p}(y_{1}) \prod_{t=1}^T 	\widehat{p}^{\mathcal{A}_t }(y_{t}|y_{1:t-1})
	\end{equation}
	with $	\widehat{p}^{\mathcal{A}_t }(y_{t}|y_{1:t-1})= \frac{1}{N}\sum_{n=1}^{N} w^{(n)}_t$ is an unbiased estimator of $p(y_{1:T})$, i.e.:
	\begin{equation}
		\mathbb{E}_{\mathcal{A}_{1:T}} \left[ \widehat{p}(y_{1:T})\right]  = p(y_{1:T}).
	\end{equation}
\end{theorem}
In order to prove Theorem \ref{theo:unbiased}, Lemma \ref{lem:Atm1} and Lemma \ref{lem:Atmhm1} below, must also hold.

\begin{lemma}\label{lem:Atm1}
	\begin{equation}
		\mathbb{E}\left[ \widehat{p}(y_t|y_{1:(t-1)})\bigg|\mathcal{A}_{t-1}\right] = \sum_{l=1}^{N} p(y_t| x_{t-1}^{(l)})\widetilde{w}^{(l)}_{t-1} .
	\end{equation}
\end{lemma}

\begin{lemma}\label{lem:Atmhm1}
	\begin{equation}\label{eq:lemma3}
		\mathbb{E}\left[ \widehat{p}(y_{(t-h):t}|y_{1:(t-h-1)})\bigg|\mathcal{A}_{t-h-1}\right] = \sum_{l=1}^{N} p(y_{(t-h):t}| x_{t-h-1}^{(l)})\widetilde{w}^{(l)}_{t-h-1} 
	\end{equation}
\end{lemma}

\subsection*{Proof of Lemma \ref{lem:Atm1}}
\begin{proof}
\begin{equation*}
	\hspace*{-0cm}\begin{split}
		\mathbb{E}\left[ \widehat{p} (y_t | y_{1:(t-1)}) \right.&\left.\bigg| \mathcal{A}_{t-1}\right] = \mathbb{E}\left[  \frac{1}{N}\sum_{n=1}^{N} w^{(n)} \bigg| \mathcal{A}_{t-1}\right]\\
		& \text{by weight definition in Equation \eqref{eq:lbweights}} \\
		& = \mathbb{E}\left[  \frac{1}{\cancel{N}} \left\lbrace \sum_{n=1}^{N-1}\frac{p\left( x_t,y_t|x_{t-1}^{(n)}\right) }{q\left( x_t|x_{t-1}^{(n)}\right) }\cdot\frac{1-\widetilde{w}_{t-1}^{(N)}r }{N-1} \cdot\cancel{ N }+\right.\right.\\
		&\left.\left. \qquad\qquad
		\frac{p\left( x_t,y_t|x_{t-1}^{(N)}\right) }{q\left( x_t|x_{t-1}^{(N)}\right)}\cdot	\widetilde{w}_{t-1}^{(N)}\cdot r\cdot \cancel{ N } 
		\right\rbrace \right]\\
		&\text{by linearity of expectation}\\
		&= \left\lbrace \sum_{n=1}^{N-1} \mathbb{E}\left[ \frac{p\left( x_t,y_t|x_{t-1}^{(n)}\right) }{q\left( x_t|x_{t-1}^{(n)}\right) }\cdot\frac{1-\widetilde{w}_{t-1}^{(N)}r }{N-1}\right] + \mathbb{E}\left[ 
		\frac{p\left( x_t,y_t|x_{t-1}^{(N)}\right) }{q\left( x_t|x_{t-1}^{(N)}\right)}\cdot\widetilde{w}_{t-1}^{(N)}\cdot r\right]
		\right\rbrace \\
	\end{split}
\end{equation*}
solve the expectation w.r.t. all the values of the resampling index $n$ (first two lines) and w.r.t. the r.v. $x_t$, proposed according to $q$ (third line).
\begin{equation*}
	\begin{split}
		&= (N-1)\cdot \sum_{x_t \in \Omega_t}^{ }\left[  \sum_{k=1}^{N-1} \frac{p\left( x_t,y_t|x_{t-1}^{(k)}\right) }{q\left( x_t|x_{t-1}^{(k)}\right) }\cdot\left(\frac{1-\widetilde{w}_{t-1}^{(N)}r }{N-1}\right)\cdot q\left(x_t|x_{t-1}^{(k)}\right)\cdot \frac{\widetilde{w}_t^{(k)}}{1-\widetilde{w}^{(N)}r} + \right.\\
		& \left. \qquad\qquad+  \frac{p\left( x_t,y_t|x_{t-1}^{(N)}\right) }{q\left( x_t|x_{t-1}^{(N)}\right) }\cdot\left(\frac{1-\widetilde{w}_{t-1}^{(N)}r }{N-1}\right)\cdot q\left(x_t|x_{t-1}^{(N)}\right)\cdot \frac{\widetilde{w}_t^{(N)} (1-r)}{1-\widetilde{w}^{(N)}r}  \right] +\\
		& \qquad + \sum_{x_t \in \Omega_t}^{ } \left[ \underbrace{\sum_{n=1}^{N-1} \frac{p\left( x_t,y_t|x_{t-1}^{(n)}\right) }{q\left( x_t|x_{t-1}^{(n)}\right) } \cdot \widetilde{w}_{t-1}^{(n)}\cdot r \cdot q\left(x_t|x_{t-1}^{(n)}\right)\cdot0}_{=0} +\frac{p\left( x_t,y_t|x_{t-1}^{(N)}\right) }{q\left( x_t|x_{t-1}^{(N)}\right) } \cdot \widetilde{w}_{t-1}^{(N)}\cdot r \cdot q\left(x_t|x_{t-1}^{(N)}\right)\cdot1 \right]\\
	\end{split}
\end{equation*}
cancel elements\begin{equation*}
\begin{split}
		&= \cancel{(N-1)}\cdot \sum_{x_t \in \Omega_t}^{ }\left[  \sum_{k=1}^{N-1} \frac{p\left( x_t,y_t|x_{t-1}^{(k)}\right) }{\cancel{q\left( x_t|x_{t-1}^{(k)}\right)} }\cdot\left(\frac{\bcancel{1-\widetilde{w}_{t-1}^{(N)}r} }{\cancel{N-1}}\right)\cdot \cancel{q\left(x_t|x_{t-1}^{(k)}\right)}\cdot \frac{\widetilde{w}_t^{(k)}}{\bcancel{1-\widetilde{w}^{(N)}r}} + \right.\\
		& \left. \qquad\qquad+  \frac{p\left( x_t,y_t|x_{t-1}^{(N)}\right) }{\cancel{q\left( x_t|x_{t-1}^{(N)}\right)} }\cdot\left(\frac{\bcancel{1-\widetilde{w}_{t-1}^{(N)}r }}{\cancel{N-1}}\right)\cdot \cancel{q\left(x_t|x_{t-1}^{(N)}\right)}\cdot \frac{\widetilde{w}_t^{(N)} (1-r)}{\bcancel{1-\widetilde{w}^{(N)}r}}  \right] +\\
		& \qquad + \sum_{x_t \in \Omega_t}^{ } \left[ \frac{p\left( x_t,y_t|x_{t-1}^{(N)}\right) }{\cancel{q\left( x_t|x_{t-1}^{(N)}\right)} } \cdot \widetilde{w}_{t-1}^{(N)}\cdot r \cdot\cancel{ q\left(x_t|x_{t-1}^{(N)}\right)}\cdot1 \right]\\
	\end{split}
\end{equation*}\begin{equation*}
\begin{split}
		&=  \sum_{x_t \in \Omega_t}^{ }\left[  \sum_{k=1}^{N-1} {p\left( x_t,y_t|x_{t-1}^{(k)}\right) }\cdot {\widetilde{w}_t^{(k)}} +   {p\left( x_t,y_t|x_{t-1}^{(N)}\right) }\cdot {\widetilde{w}_t^{(N)} (1-r)}  \right] +\\
		& \qquad \qquad+ \sum_{x_t \in \Omega_t}^{ } \left[ {p\left( x_t,y_t|x_{t-1}^{(N)}\right) }\cdot \widetilde{w}_{t-1}^{(N)}\cdot r  \right]\\		\end{split}
	\end{equation*}\begin{equation*}
	\begin{split}
		&\text{split summation inside brackets}\\
		&=  \sum_{x_t \in \Omega_t}^{ }\left[  \sum_{k=1}^{N-1} {p\left( x_t,y_t|x_{t-1}^{(k)}\right) }\cdot {\widetilde{w}_t^{(k)}}\right] +   (1-r)\sum_{x_t \in \Omega_t}^{ }\left[   {p\left( x_t,y_t|x_{t-1}^{(N)}\right) }\cdot {\widetilde{w}_t^{(N)} }  \right] +\\
	&	\qquad\qquad +r\sum_{x_t \in \Omega_t}^{ } \left[ {p\left( x_t,y_t| x_{t-1}^{(N)}\right) }\cdot \widetilde{w}_{t-1}^{(N)}  \right]\\
		&=  \sum_{x_t \in \Omega_t}^{ }\left[  \sum_{k=1}^{N} {p\left( x_t,y_t|x_{t-1}^{(k)}\right) }\cdot {\widetilde{w}_t^{(k)}}\right]\\
		&\text{swap summations}\\
		&= \sum_{k=1}^{N} \left[  \sum_{x_t \in \Omega_t}^{ } {p\left( x_t,y_t|x_{t-1}^{(k)}\right) }\cdot {\widetilde{w}_t^{(k)}}\right]= \sum_{k=1}^{N} \left[ {p\left(y_t|x_{t-1}^{(k)}\right) }\cdot {\widetilde{w}_t^{(k)}}\right]  \\
	\end{split}
\end{equation*}
\end{proof}

\subsection*{Proof of Lemma \ref{lem:Atmhm1}}
This proof is by induction.
\begin{proof}
	\textit{Initialisation}: for $h=0$ Lemma \ref{lem:Atmhm1} reduces to Lemma \ref{lem:Atm1} and hence holds. 
	\begin{equation*}
		\mathbb{E}\left[ \widehat{p}(y_{t}|y_{1:(t-1)})\bigg|\mathcal{A}_{t-1}\right] = \sum_{l=1}^{N} p(y_{:t}| x_{t-1}^{(l)})\widetilde{w}^{(l)}_{t-1} 
	\end{equation*}
	\textit{Inductive hypothesis}: Lemma \ref{lem:Atmhm1} holds for a general $h$. \\
	The following inductive steps prove that if Lemma \ref{lem:Atmhm1} holds for a general $h$, it also holds for $h+1$. \\
	Assume \eqref{eq:lemma3} and compute $ \mathbb{E}\left[ \widehat{p}(y_{(t-l):t}|y_{1:(t-l-1)})\bigg|\mathcal{A}_{t-l-1}\right]$ for $l=h+1$. 
	
	\begin{equation*}
		\begin{split}
			\mathbb{E}\left[ \widehat{p}(y_{(t-h-1):t}|\right.&\left.y_{1:(t-h-2)})\bigg|\mathcal{A}_{t-h-2}\right]= \mathbb{E}\left[ \widehat{p}(y_{(t-h-1)}, y_{(t-h):t}|y_{1:(t-h-2)})\bigg|\mathcal{A}_{t-h-2}\right]\\
			&= \mathbb{E}\left[ \left\lbrace\widehat{p}(y_{(t-h):t}|y_{(t-h-1)}, y_{1:(t-h-2)}) \widehat{p}(y_{(t-h-1)}|y_{1:(t-h-2)})\right\rbrace\bigg|\mathcal{A}_{t-h-2}\right]\\
			&= \mathbb{E}\left[ \left\lbrace\widehat{p}(y_{(t-h):t}|y_{(t-h-1)}) \widehat{p}(y_{(t-h-1)}|y_{1:(t-h-2)})\right\rbrace\bigg|\mathcal{A}_{t-h-2}\right]\\
			&\text{from the law of total expectation $\mathbb{E}(X)=\mathbb{E}(\mathbb{E}(X|Y))$, we can write:}\\
			&= \mathbb{E}\left[  \mathbb{E}\left[ \left\lbrace\widehat{p}(y_{(t-h):t}|y_{(t-h-1)}) \underbrace{\widehat{p}(y_{(t-h-1)}|y_{1:(t-h-2)})}_{\text{constant conditionally on}\mathcal{A}_{t-h-1}}\right\rbrace\bigg|\mathcal{A}_{t-h-1}\right]\bigg|\mathcal{A}_{t-h-2}\right]\\	
			&= \mathbb{E}\left[  \underbrace{\mathbb{E}\left[ \widehat{p}(y_{(t-h):t}|y_{(t-h-1)}) \bigg|\mathcal{A}_{t-h-1}\right]}_{\text{given in initialisation}\eqref{eq:lemma3}}\widehat{p}(y_{(t-h-1)}|y_{1:(t-h-2)})\bigg|\mathcal{A}_{t-h-2}\right]\\
				\end{split}
		\end{equation*}
	\begin{equation*}
	\begin{split}&= \mathbb{E}\left[ \left[  \sum_{k=1}^{N} p(y_{(t-h):t}| x_{t-h-1}^{(k)})\widetilde{w}^{(k)}_{t-h-1} \right] \underbrace{\widehat{p}(y_{(t-h-1)}|y_{1:(t-h-2)})}_{\text{by def. of }\widehat{p}(y_t|y_{1:(t-1)}) \text{ with } t=t-h }\bigg|\mathcal{A}_{t-h-2}\right]\\
			&= \mathbb{E}\left[ \left[  \sum_{k=1}^{N} p(y_{(t-h):t}| x_{t-h-1}^{(k)})\widetilde{w}^{(k)}_{t-h-1} \right] \frac{1}{N} \sum_{k=1}^Nw_{t-h-1}^{(k)} \bigg|\mathcal{A}_{t-h-2}\right]\\
			& \text{from definition of self-normalised weights we have}\\
			&= \mathbb{E}\left[ \left[  \sum_{k=1}^{N} p(y_{(t-h):t}| x_{t-h-1}^{(k)})\frac{{w}^{(k)}_{t-h-1}}{\cancel{\sum_{l=1}^{N}{w}^{(l)}_{t-h-1}}} \right] \frac{1}{N} \cancel{\sum_{k=1}^Nw_{t-h-1}^{(k)}} \bigg|\mathcal{A}_{t-h-2}\right]\\
			&\text{from linearity of expectation we have}\\
			&= \frac{1}{N} \sum_{k=1}^{N}\mathbb{E}\left[ \left[  p(y_{(t-h):t}| x_{t-h-1}^{(k)}){{w}^{(k)}_{t-h-1}} \right]  \bigg|\mathcal{A}_{t-h-2}\right]\\
			&= \frac{1}{N} \sum_{k=1}^{N}\mathbb{E}\left[ \left[  p(y_{(t-h):t}| x_{t-h-1}^{(k)}){{w}^{(k)}_{t-h-1}} \right]  \bigg|\mathcal{A}_{t-h-2}\right]\\
			&\text{separate the $N$-th element of the summation and use linearity of expectation}\\
			&= \frac{1}{N} \sum_{k=1}^{N-1}\mathbb{E}\left[ \left[  p(y_{(t-h):t}| x_{t-h-1}^{(k)}){{w}^{(k)}_{t-h-1}} \right]  \bigg|\mathcal{A}_{t-h-2}\right] + \mathbb{E}\left[ \left[  p(y_{(t-h):t}| x_{t-h-1}^{(N)}){{w}^{(N)}_{t-h-1}} \right]  \bigg|\mathcal{A}_{t-h-2}\right]\\
			&\text{write the weights explicitly according to \eqref{eq:lbweights}}\\
			&=\cancel{ \frac{1}{N}} \sum_{k=1}^{N-1}\mathbb{E}\left[ \left[  p(y_{(t-h):t}| x_{t-h-1}^{(k)}) \frac{p\left( x_{t-h-1},y_{t-h-1}|x_{t-h-2}^{(a_k)}\right) }{q\left( x_{t-h-1}|x_{t-h-2}^{(a_k)}\right) }\cdot\frac{1-\widetilde{w}_{t-h-2}^{(N)}\cdot r }{N-1} \cdot \cancel{N} \right]  \bigg|\mathcal{A}_{t-h-2}\right] +\\
			&\qquad\qquad  \mathbb{E}\left[ \left[  p(y_{(t-h):t}| x_{t-h-1}^{(N)}) \frac{p\left( x_{t-h-1},y_{t-h-1}|x_{t-h-2}^{(a_k)}\right) }{q\left( x_{t-h-1}|x_{t-h-2}^{(a_k)}\right) }\cdot{\widetilde{w}_{t-h-2}^{(N)}\cdot r \cdot \cancel{N}} \right]  \bigg|\mathcal{A}_{t-h-2}\right]\\
			&\text{write explicitly the expectation w.r.t. $x_{t-h-1}$ and the resampling index $a_k$}\\
			&= {(N-1)}\left[ \sum_{l=1}^{N-1} \sum_{x_{t-h-1}\in\Omega}^{} \frac{p\left(y_{(t-h):t}| x_{t-h-1}^{(k)}\right) p\left( x_{t-h-1},y_{t-h-1}|x_{t-h-2}^{(l)}\right) }{q\left( x_{t-h-1}|x_{t-h-2}^{(l)}\right) }\cdot\frac{1-\widetilde{w}_{t-h-2}^{(N)}\cdot r }{N-1} \cdot \right.\\
			&\qquad \qquad \qquad q\left( x_{t-h-1}|x_{t-h-2}^{(l)}\right)\cdot\frac{\widetilde{w}^{(l)}_{t-h-1}}{1-\widetilde{w}_{t-h-2}^{(N)}\cdot r }  + \sum_{x_{t-h-1}\in\Omega}^{} \frac{p\left(y_{(t-h):t}| x_{t-h-1}^{(k)}\right) p\left( x_{t-h-1},y_{t-h-1}|x_{t-h-2}^{(N)}\right) }{q\left( x_{t-h-1}|x_{t-h-2}^{(N)}\right) } \cdot\\
			&\qquad\qquad\qquad \left.
			\frac{1-\widetilde{w}_{t-h-2}^{(N)}\cdot r }{N-1} \cdot q\left( x_{t-h-1}|x_{t-h-2}^{(N)}\right)\cdot\frac{\widetilde{w}^{(N)}_{t-h-1}(1-r)}{1-\widetilde{w}_{t-h-2}^{(N)}\cdot r }  \right]  +\\
			&  \sum_{x_{t-h-1}\in\Omega}^{} \frac{p\left(y_{(t-h):t}| x_{t-h-1}^{(k)}\right) p\left( x_{t-h-1},y_{t-h-1}|x_{t-h-2}^{(N)}\right) }{q\left( x_{t-h-1}|x_{t-h-2}^{(N)}\right) }\cdot{\widetilde{w}_{t-h-2}^{(N)}\cdot r } \cdot q\left( x_{t-h-1}|x_{t-h-2}^{(N)}\right)\cdot1 \\
			&\text{which allows for the following simplification}\\
			\end{split}
		\end{equation*}
	\begin{equation*}
	\begin{split}
			&= \cancel{(N-1)}\left[ \sum_{l=1}^{N-1} \sum_{x_{t-h-1}\in\Omega}^{} \frac{p\left(y_{(t-h):t}| x_{t-h-1}^{(k)}\right) p\left( x_{t-h-1},y_{t-h-1}|x_{t-h-2}^{(l)}\right) }{\bcancel{q\left( x_{t-h-1}|x_{t-h-2}^{(l)}\right)} }\cdot\right.\\
			&\qquad \qquad \qquad\frac{\cancel{1-\widetilde{w}_{t-h-2}^{(N)}\cdot r }}{\cancel{N-1}} \cdot \bcancel{ q\left( x_{t-h-1}|x_{t-h-2}^{(l)}\right)}\cdot\frac{\widetilde{w}^{(l)}_{t-h-1}}{\cancel{1-\widetilde{w}_{t-h-2}^{(N)}\cdot r} }  + \\
			&+\sum_{x_{t-h-1}\in\Omega}^{} \frac{p\left(y_{(t-h):t}| x_{t-h-1}^{(k)}\right) p\left( x_{t-h-1},y_{t-h-1}|x_{t-h-2}^{(N)}\right) }{\bcancel{q\left( x_{t-h-1}|x_{t-h-2}^{(N)}\right) }} \cdot\\
			&\qquad\qquad\qquad \left.
			\frac{\cancel{1-\widetilde{w}_{t-h-2}^{(N)}\cdot r }}{\cancel{N-1}} \cdot \bcancel{q\left( x_{t-h-1}|x_{t-h-2}^{(N)}\right)}\cdot\frac{\widetilde{w}^{(N)}_{t-h-1}(1-r)}{\cancel{1-\widetilde{w}_{t-h-2}^{(N)}\cdot r} }  \right]  +\\
			&  \sum_{x_{t-h-1}\in\Omega}^{} \frac{p\left(y_{(t-h):t}| x_{t-h-1}^{(k)}\right) p\left( x_{t-h-1},y_{t-h-1}|x_{t-h-2}^{(N)}\right) }{\bcancel{q\left( x_{t-h-1}|x_{t-h-2}^{(N)}\right) }}\cdot{\widetilde{w}_{t-h-2}^{(N)}\cdot r } \cdot \bcancel{q\left( x_{t-h-1}|x_{t-h-2}^{(N)}\right)}\cdot1 \\
			&=\sum_{l=1}^{N-1} \sum_{x_{t-h-1}\in\Omega}^{}p\left(y_{(t-h):t}| x_{t-h-1}^{(k)}\right) p\left( x_{t-h-1},y_{t-h-1}|x_{t-h-2}^{(l)}\right) \cdot\widetilde{w}^{(l)}_{t-h-1}+\\
			&\qquad \sum_{x_{t-h-1}\in\Omega}^{}p\left(y_{(t-h):t}| x_{t-h-1}^{(k)}\right) p\left( x_{t-h-1},y_{t-h-1}|x_{t-h-2}^{(N)}\right) \widetilde{w}^{(N)}_{t-h-1}(1-r)+\\
			&\qquad \sum_{x_{t-h-1}\in\Omega}^{} p\left(y_{(t-h):t}| x_{t-h-1}^{(k)}\right) p\left( x_{t-h-1},y_{t-h-1}|x_{t-h-2}^{(N)}\right)\cdot{\widetilde{w}_{t-h-2}^{(N)}\cdot r }\\
			&=\sum_{l=1}^{N} \sum_{x_{t-h-1}\in\Omega}^{}p\left(y_{(t-h):t}| x_{t-h-1}^{(k)}\right) p\left( x_{t-h-1},y_{t-h-1}|x_{t-h-2}^{(l)}\right) \cdot\widetilde{w}^{(l)}_{t-h-1}\\
			&=\sum_{l=1}^{N}\widetilde{w}^{(l)}_{t-h-1} \sum_{x_{t-h-1}\in\Omega}^{}p\left(y_{(t-h):t}| x_{t-h-1}^{(k)}\right) p\left( y_{t-h-1}|x_{t-h-1}^{(l)}\right) p\left( x_{t-h-1}|x_{t-h-2}^{(l)}\right) \\
			&\text{from the law of total probabilities}\\
			&=\sum_{l=1}^{N}\widetilde{w}^{(l)}_{t-h-1} p\left(y_{(t-h):t}| x_{t-h-2}^{(l)}\right) \\
		\end{split}
	\end{equation*}
	As Lemma \ref{lem:Atmhm1} holds for $h=0$, it holds for all $h$s by induction. \end{proof}
\subsection*{Proof of Theorem \ref{theo:unbiased}}
\begin{proof}
	
	From Lemma \ref{lem:Atmhm1} with $h=t-1$ we have that:
	\begin{equation*}
		\mathbb{E}\left[ \widehat{p} (y_{1:T})|\mathcal{A}_0\right] =\sum_{n=1}^{N}p(y_{1:T}|x_0^{(n)}) \widetilde{w}_0^{(n)}
	\end{equation*}
	with $ x_0^{(n)}\sim p(x_0) $ and $\widetilde{w}_0^{(n)}=\frac{1 }{N}$.
	\begin{equation}
		\mathbb{E} \left[\sum_{n=1}^{N}p(y_{1:T}|x_0^{(n)})\widetilde{w}_0^{(n)} \right]  = N \sum_{x_{0}\in\Omega}^{} p(y_{1:T}|x_0^{(n)})\cdot \frac{1}{N}\cdot p(x_0)=p(y_{1:T})
	\end{equation}\end{proof}

\printbibliography

\begin{acronym}
	\acro{CFR}{case fatality risk}
	\acro{hCFR}{hospitalised case fatality risk}
	%	\acro{iCFR}{intensive-care case fatality risk}
	\acro{sCFR}{symptomatic case fatality risk}
	%	\acro{IFR}{infection fatality risk}
	\acro{iid}{independent and identically distributed}
	\acro{SMC}{sequential Monte Carlo}
	\acro{SSM}{state-space model}
	\acro{ESS}{effective sample size}
	%	\acro{MSM}{multi-state model}
	%	\acro{ML}{maximum likelihood}
	%	\acro{HMM}{hidden Markov model}
	%	\acro{POMP}{partially observed Markov process}
	\acro{r.v.}{random variable}
	\acro{MH}{Metropolis Hastings}
	\acro{pMCMC}{particle Markov chain Monte Carlo}
	\acro{MCMC}{Markov chain Monte Carlo}
	\acro{MC}{Monte Carlo}
	\acro{USISS}{UK severe influenza surveillance system}
	%	\acro{PHE}{Public Health England}
	%	\acro{NHS}{National Health Service}
	%	\acro{ICU}{Intensive Care Unit}
	%	\acro{HDU}{High Dependence Unit}
	\acro{WHO}{World Health Organization}
	\acro{ABC}{approximate Bayesian computation}
	%	\acro{CrI}{credible interval}
	\acro{DAG}{directed acyclic graph}
	\acro{BPF}{bootstrap particle filter}
	\acro{GIMH}{grouped independence Metropolis Hastings}
	\acro{PMMH}{particle marginal Metropolis Hastings}
	\acro{MCWM}{Monte Carlo within Metropolis}
	%	\acro{CI}{confidence interval}
	%	\acro{HMC}{Hamiltonian Monte Carlo}
	\acro{APF}{alive particle filter}
	\acro{AuPF}{auxiliary particle filter}
	\acro{NPF}{nudged particle filter}
	\acro{RRS}{regional re-sampling}
	\acro{LBPF}{lifebelt particle filter}
	\acro{SIS}{sequential importance sampling}
	\acro{SIRS}{sequential importance sampling with resampling}
	\acro{p.d.f.}{probability density function}
	\acro{DMIS}{deterministic mixture importance sampling}
	\acro{MIS}{multiple importance sampling}
	\acrodefplural{r.v.}[r.v.'s]{random varaibles}
\end{acronym}

\medskip
% The information below will be filled in by AIMS production staff.
Received xxxx 20xx; revised xxxx 20xx; early access xxxx 20xx.
\medskip

\end{document}